\documentclass[twocolumn,10pt]{article}

\usepackage[utf8]{inputenc}
\usepackage[T1]{fontenc}
\usepackage{lmodern}
\usepackage[english]{babel}
\usepackage{amsmath,amssymb,amsfonts}
\usepackage{graphicx}
\usepackage{subcaption}
\usepackage{booktabs}
\usepackage{array}
\newcolumntype{C}[1]{>{\centering\arraybackslash}m{#1}}
\usepackage{multirow}
\usepackage{longtable}
\usepackage{xcolor}
\usepackage{hyperref}
\hypersetup{
    colorlinks=true,
    linkcolor=blue!70!black,
    citecolor=blue!70!black,
    urlcolor=blue!70!black
}
\usepackage[numbers,sort&compress]{natbib}
\usepackage{geometry}
\usepackage{microtype}
\usepackage{authblk}
\usepackage{titlesec}
\usepackage{abstract}
\usepackage{fancyhdr}
\usepackage{caption}
\titleformat{\section}{\normalfont\large\bfseries}{\thesection.}{0.5em}{}
\titleformat{\subsection}{\normalfont\normalsize\bfseries}{\thesubsection.}{0.5em}{}
\titleformat{\subsubsection}{\normalfont\normalsize\itshape}{\thesubsubsection.}{0.5em}{}

\fancypagestyle{plain}{\fancyhf{}\fancyfoot[C]{\thepage}}

\title{\textbf{Toward Scalable Heterogeneous Quantum Networks:\\
Microwave-Optical Transduction Across Platforms}}

\author{
  Tarvir Anjum Aditto$^{1,2,*}$,\,
  Jaiyan Sadid Ifty$^{1,*}$,\,
  and Khondokar Zahin$^{1,*}$\\
  {\small $^*$These authors contributed equally to this work.}\\
  {\small $^1$Department of Electrical and Electronic Engineering, BUET, Dhaka, Bangladesh}\\
  {\small $^2$Department of Computer Science and Engineering, BRAC University, Dhaka, Bangladesh}
}
\date{}

\begin{document}

\maketitle
\thispagestyle{plain}

\begin{abstract}
The development of scalable quantum networks requires coherent interfaces capable of converting microwave photons used in superconducting quantum processors into optical photons suitable for long-distance fiber transmission.
This review surveys recent progress in microwave-to-optical quantum transduction across optomechanical, electro-optic, and magneto-optic platforms, with emphasis on conversion efficiency, bandwidth, added noise, and operating temperature. In
addition to standard metrics, we propose the internal efficiency
$\eta_{\mathrm{in}}$ and the magnon decay rate $\kappa_m/2\pi$ as normalized parameters that enable fairer comparison across heterogeneous implementations. Optomechanical systems achieve internal phonon-to-photon efficiencies of 93\%
with sub-quantum added noise of 0.25 quanta at millikelvin temperatures. Electro-optic devices based on LiNbO$_3$ and AlN have advanced from room-temperature efficiencies below 1\% to millikelvin systems with internal efficiencies approaching 99.5\%, added noise as low as 0.16 quanta at 60~mK,
and bandwidths extending to several tens of megahertz. Magneto-optic
(optomagnonic) platforms exhibit the lowest efficiencies (typically $10^{-10}$ to $10^{-8}$), but offer intrinsic non-reciprocity and broadband magnonic operation, with emerging approaches based on topological heterostructures and
magnon squeezing predicting enhancements up to $10^{-4}$. Optomechanical systems appear promising for high-fidelity quantum state transfer, electro-optic transducers for high-bandwidth coherent links, and magneto-optic
devices for non-reciprocal network components. We discuss the fundamental trade-off between efficiency and added noise across all three platforms, and argue that heterogeneous microwave-optical transduction is emerging as a key enabling
technology for distributed quantum computing and large-scale quantum networks.
\end{abstract}

\vspace{4pt}

\section{Introduction}

Transduction refers to the conversion of a signal from one physical domain or frequency to another. In quantum technologies, microwave-to-optical quantum transduction is especially important because superconducting qubits, one of the leading platforms for quantum computation, operate in the microwave frequency range (1--10~GHz) and require millikelvin temperatures to preserve quantum
coherence~\cite{Devoret2013,Blais2021,QTe}. These systems support fast gate operations and benefit from mature integrated-circuit fabrication compatibility, yet face significant scaling limits: a single dilution refrigerator can host only a limited number of qubits, and thermal load rises rapidly with system
complexity~\cite{Sekine2025Review}. Fault-tolerant quantum computing further compounds this challenge, since encoding one logical qubit requires many physical qubits~\cite{Ref3_Gottesman2010}, motivating distributed quantum computing architectures that interconnect multiple cryogenic nodes.

\begin{figure*}[h!]
    \centering
    \includegraphics[width=0.49\textwidth]{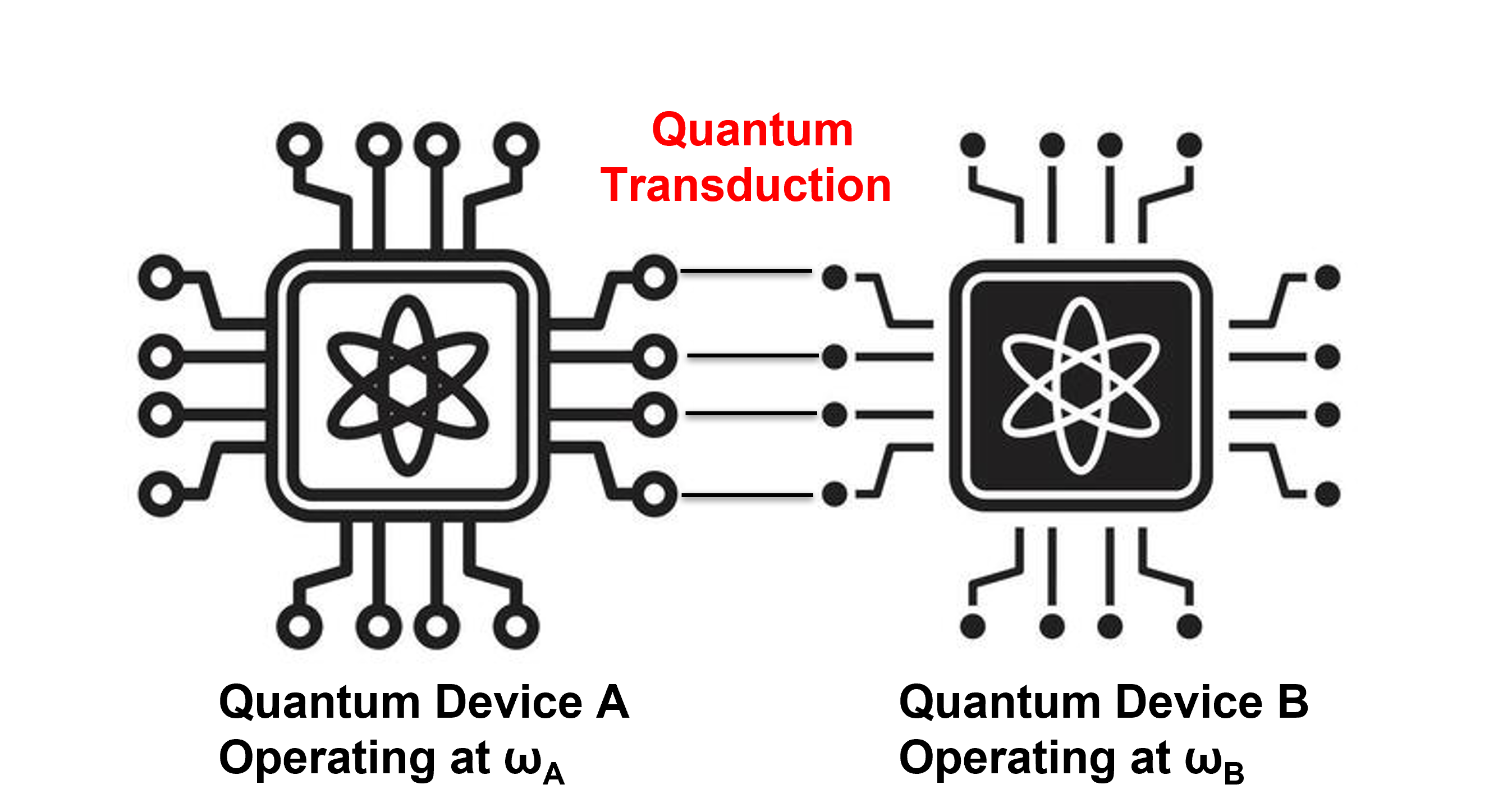}
    \hfill
    \includegraphics[width=0.49\textwidth]{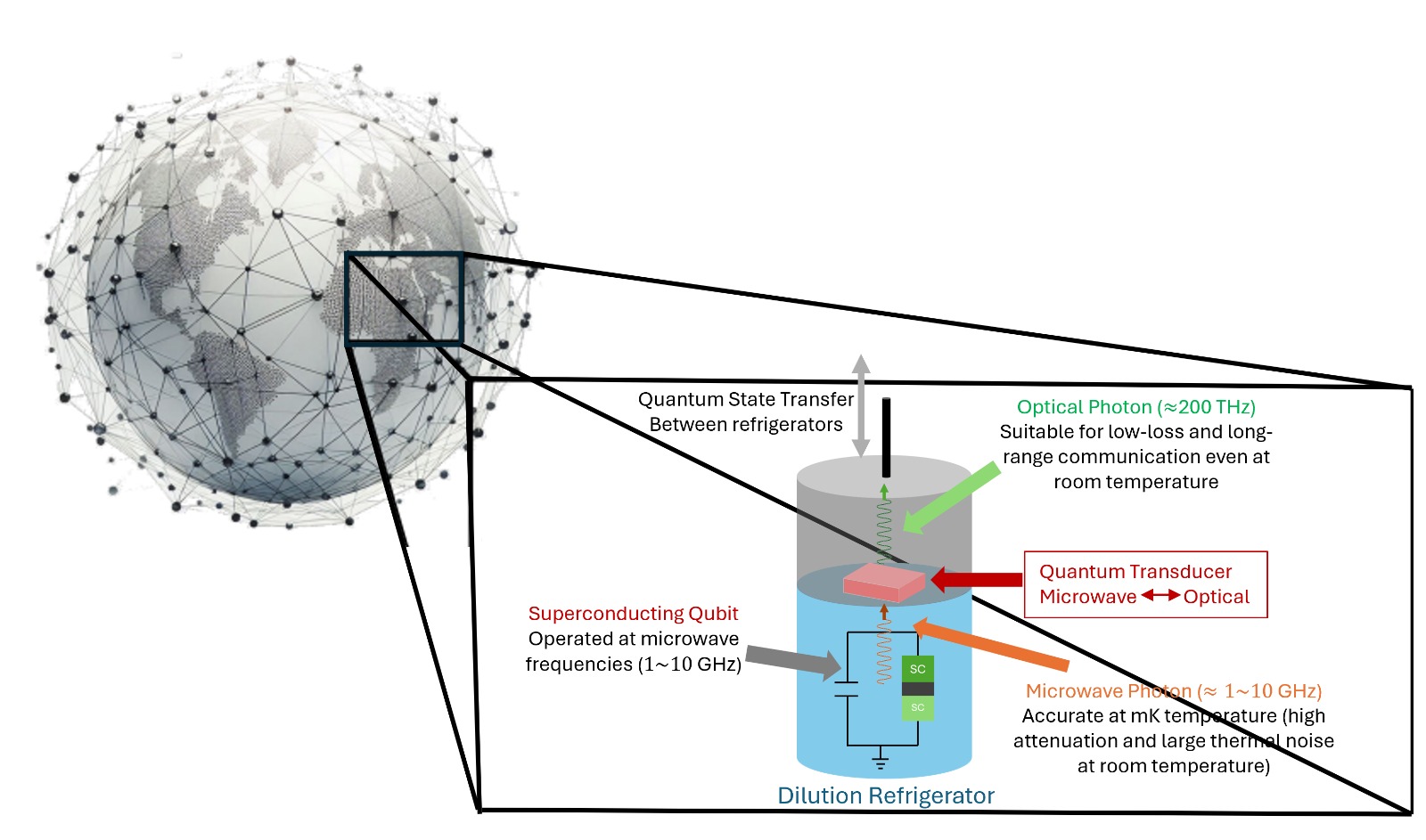}
    \caption{Quantum transduction concept.
    \textbf{Left}: Schematic representation of two quantum devices operating at
    different frequencies connected via a quantum transduction interface.
    \textbf{Right}: Illustration of microwave-to-optical conversion inside a
    dilution refrigerator enabling long-distance quantum state transfer.}
    \label{fig:quantum_transduction}
\end{figure*}

Microwave photons, however, perform poorly in long-distance links. At room temperature they experience strong attenuation, are highly susceptible to thermal noise, and cryogenic waveguide alternatives are costly, complex, and difficult to scale~\cite{Ref4_Cleland2003}. Optical photons, by contrast, exhibit extremely low fiber attenuation (approximately 0.2~dB/km) and are robust against room-temperature noise~\cite{Ref5_Kimble2008}, though they do
not interact strongly enough with each other to serve directly as qubit carriers~\cite{qte22}. Microwave-optical quantum transduction provides a path to bridge these two regimes, enabling communication between superconducting processors and optical quantum networks. Fig.~\ref{fig:quantum_transduction} illustrates an overview of the transduction process. This interface is essential because no single hardware platform currently satisfies the combined requirements of computation, memory, and long-distance communication~\cite{Ref5_Kimble2008,
Ref7_Awschalom2021}, and interconnecting complementary platforms offers a concrete route toward a scalable \textit{quantum internet}.

The technical challenges are substantial. The frequency gap between the microwave and optical domains spans nearly five orders of magnitude, making direct conversion non-trivial. Beyond the engineering difficulty, quantum transduction is fundamentally more constrained than its classical counterpart: the no-cloning theorem prohibits copying or amplifying an arbitrary quantum
state~\cite{Ref8_Wootters1982}, and any intermediate measurement collapses the quantum state being transferred~\cite{NielsenChuang2000}. These constraints
rule out the regeneration and amplification strategies available in classical frequency conversion, requiring instead coherent physical interactions that preserve the full quantum character of the signal throughout.
To address these challenges, three main physical approaches have been explored: optomechanical, electro-optic, and magneto-optic interactions. These platforms enable microwave-optical coupling via single-stage or multi-stage configurations
through intermediate bosonic modes or direct nonlinear
interactions~\cite{Ref6_Lauk2020,Ref9_Fong2020}. In this review, we examine the progress of microwave-to-optical transduction across these platforms. We summarize recent experimental advances and compare key performance metrics, including efficiency, bandwidth, cooperativity, and added noise. In cases where a metric is not explicitly reported in the original work, we derive it from the published experimental data, enabling a consistent comparison across all three platforms. Beyond these standard measures, we propose the internal efficiency $\eta_{\mathrm{in}}$ and
the magnon decay rate $\kappa_m/2\pi$ as normalized parameters that are not consistently defined in the existing literature but prove important for fair comparison across heterogeneous implementations. Finally, we analyze system-level trade-offs, deployment challenges, and application-specific platform selection. Together, the proposed metrics and cross-platform analysis provide a foundation for evaluating and guiding the next generation of microwave-optical quantum transduction experiments.

\section{Scope}

This review covers experimental and theoretical progress in
microwave-to-optical quantum transduction from 2014 to 2026, focusing on three main physical platforms: optomechanical, electro-optic, and magneto-optic systems. The primary focus is on performance metrics relevant to quantum
networking applications, specifically conversion efficiency, added noise, cooperativity, transduction bandwidth, and operating temperature.  Sources include peer-reviewed journal articles, conference proceedings, and preprints;
foundational textbooks are cited where needed to establish the theoretical framework. This review does not cover rare-earth ion systems, atomic or molecular transduction schemes, or purely classical microwave-optical conversion. 

\section{General Considerations for Quantum Transduction}
\begin{figure*}[h!]
    \centering
    \includegraphics[width=\linewidth]{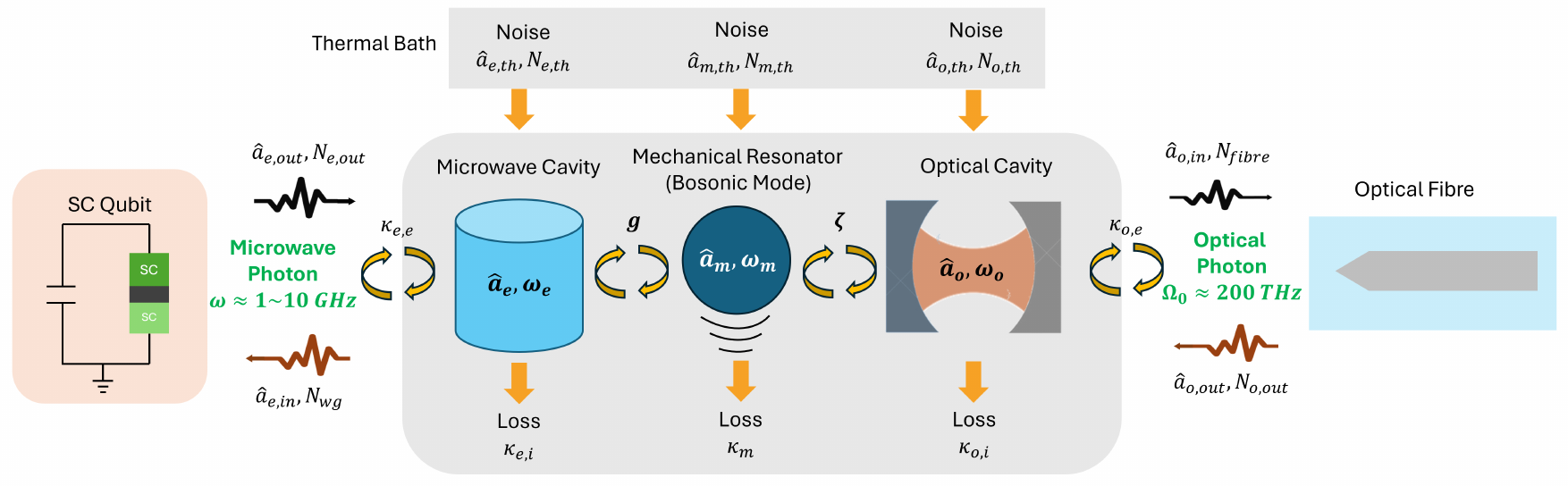}
    \caption{Schematic of the physical quantities governing one-stage quantum
    transduction. A single intermediate bosonic mode (phonon or magnon)
    coherently couples a microwave cavity to an optical cavity. The system is
    described by cavity and bath operators together with mode frequencies,
    coherent coupling rates, and intrinsic loss channels. Intrinsic losses are
    modeled as couplings to thermal baths. Optical and fiber noise contributions
    are negligible at the optical frequency of approximately 200~THz.}
    \label{fig:process}
\end{figure*}
\subsection{Superconducting Qubits and Microwave-Optical Quantum Transduction}

Superconducting quantum processors exploit the formation of Cooper pairs at
cryogenic temperatures, where lattice-mediated electron pairing produces a
macroscopic condensate carrying a coherent phase~\cite{BCS1957}. This global
superconducting phase enables Josephson tunneling across an insulating barrier,
giving rise to a nonlinear inductive element whose cosine potential supports
quantized energy levels, the basis of superconducting
qubits~\cite{Josephson1962,Devoret2013}. When embedded in microwave resonators,
these qubits coherently exchange excitations with cavity modes in the GHz
frequency range~\cite{Blais2021}.

Although superconducting qubits provide a scalable platform for quantum
information processing, microwave photons are unsuitable for long-distance
transmission owing to thermal noise and attenuation outside dilution
refrigerators. This motivates the development of microwave-optical quantum
transduction. In general, the transduction process can be described as an
$N$-stage architecture, where $N$ denotes the number of intermediate bosonic
modes~\cite{Lambert2020,Han2021}. The case $N=0$ corresponds to direct
nonlinear photon interaction via electro-optic wave mixing, while $N=1$ involves
a single intermediate oscillator such as a mechanical phonon
mode~\cite{Han2021}. Practical implementations based on optomechanical,
electro-optic, and magneto-optic platforms have demonstrated coherent conversion
between microwave and optical
photons~\cite{Bartholomew2020,Wang2022,Rochman2023}.

Within an effective input-output circuit representation, each stage of the
transducer is characterized by coherent coupling strengths and intrinsic losses. Fig.~\ref{fig:process} summarizes the physical quantities and source of losses involved in the transduction mechanism.
The conversion efficiency $\eta$ is defined as the probability that an input
microwave photon is converted into an output optical photon. Added noise
$N_{\mathrm{add}}$ is the average number of spurious photons introduced during
conversion~\cite{Lambert2020}. Achieving high efficiency requires large
cooperativity $C = 4g^{2}/(\kappa_{1}\kappa_{2})$ at each stage. For quantum
networking, operating in the quantum regime $N_{\mathrm{add}} < 1$ is more
important than maximizing $\eta$, since optical loss can be mitigated through
heralded entanglement, whereas excess noise permanently degrades
fidelity~\cite{Han2021}.

\subsection{Quantum Capacity and Conditions for Quantum Transduction}

A linear microwave-optical quantum transducer can be modeled as a bosonic
Gaussian attenuator channel, where a beam-splitter unitary couples the signal
mode $\hat{a}$ to an environmental mode $\hat{c}$. In the Heisenberg picture,
the frequency-domain input-output relation is

\begin{equation}
\hat{a}_{\mathrm{out}}(\omega)
= \sqrt{\eta(\omega)}\,\hat{a}_{\mathrm{in}}(\omega)
+ \sqrt{1-\eta(\omega)}\,\hat{c}_{\mathrm{in}}(\omega),
\label{eq:input_output}
\end{equation}

where $\eta(\omega)$ is the transduction efficiency, representing the fraction
of the input faithfully transmitted, and the remainder is lost to the
environment~\cite{Gardiner2004,Walls2008}. The quantum capacity of this channel
is determined by the coherent information $I_c = S(\rho_B) - S(\rho_E)$, where
$S$ denotes the von Neumann entropy. For a thermal input state with mean photon
number $\bar{n}$,

\begin{equation}
I_c(\eta,\bar{n}) = g(\eta\bar{n}) - g((1-\eta)\bar{n}),
\label{eq:coherent_info}
\end{equation}
where $g(x)$ is the bosonic thermal entropy function defined in
Eq.~(\ref{eq:entropy_func}).
\begin{equation}
g(x) = (x+1)\log_2(x+1) - x\log_2 x
\label{eq:entropy_func}
\end{equation}
The channel is degradable for $\eta(\omega) > 1/2$ and anti-degradable
otherwise, so the single-mode quantum capacity is

\begin{equation}
q_1(\omega) =
\begin{cases}
\log_2\!\left(\dfrac{\eta(\omega)}{1-\eta(\omega)}\right), & \eta(\omega) >
\tfrac{1}{2},\\[8pt]
0, & \eta(\omega) \leq \tfrac{1}{2}.
\end{cases}
\label{eq:q1}
\end{equation}

At finite temperature the environmental mode acquires a mean occupation
$\bar{n}(\omega) = (e^{\hbar\omega/k_BT}-1)^{-1}$, converting the pure-loss
channel into a thermal-loss Gaussian channel. Excess thermal photons can
suppress the coherent information entirely, even when $\eta > 1/2$. The
input-referred added noise is

\begin{equation}
N_{\mathrm{add}}(\omega) = (1 - \eta(\omega))\,\bar{n}(\omega).
\label{eq:nadd}
\end{equation}

Quantum-enabled operation therefore requires both conditions to hold
simultaneously~\cite{Ref6_Lauk2020,Caruso2006}:

\begin{equation}
\eta(\omega) > \frac{1}{2}
\qquad \text{and} \qquad
N_{\mathrm{add}}(\omega) \ll 1.
\label{eq:quantum_regime}
\end{equation}

Since the performance of a realistic transducer varies across frequency, the
overall quantum communication rate integrates over all spectral modes,

\begin{equation}
Q_1 = \int \frac{d\omega}{2\pi}\,q_1(\omega),
\label{eq:Q1}
\end{equation}

and only regions satisfying both conditions in Eq.~(\ref{eq:quantum_regime})
contribute positively. Achieving high efficiency, effective cryogenic
suppression of thermal photons, and adequate bandwidth are therefore jointly
essential for scalable quantum interconnects.
\section{Performance Metrics for Quantum Transduction}

We consider a single-stage quantum transduction process mediated by a single
intermediate bosonic mode, as illustrated in Fig.~\ref{fig:process}. The system
comprises a microwave cavity mode $\hat{a}_e$, an intermediate bosonic mode
$\hat{a}_m$, and an optical cavity mode $\hat{a}_o$, each coupled to
corresponding input, output, and thermal bath channels. The intracavity dynamics
obey

\begin{equation}
\dot{\vec{c}} = -A\vec{c} - B\vec{c}_{\mathrm{in}},
\end{equation}

where $\vec{c} = [\hat{a}_e,\hat{a}_m,\hat{a}_o]^T$, with total loss rates
$\kappa_e = \kappa_{e,e}+\kappa_{e,i}$ and $\kappa_o = \kappa_{o,e}+\kappa_{o,i}$.
The input-output relation
$\vec{c}_{\mathrm{out}}(\omega) = S(\omega)\vec{c}_{\mathrm{in}}(\omega)$
is governed by the scattering matrix $S(\omega)$, from which all performance
metrics are derived.

\subsection{Transduction Efficiency}

The microwave-to-optical transduction efficiency $\eta$ is defined as the
probability that an incoming microwave photon emerges as an outgoing optical
photon:

\begin{equation}
\eta(\omega) = |S_{oe}(\omega)|^{2}.
\end{equation}

Under resonance conditions this reduces to

\begin{equation}
\eta = \eta_e\eta_o\,\frac{4C_{em}C_{om}}{(1+C_{em}+C_{om})^{2}},
\label{eq:eta}
\end{equation}

where $C_{em}$ and $C_{om}$ are the electromechanical and optomechanical
cooperativities, and $\eta_e$, $\eta_o$ are the external coupling efficiencies
of the microwave and optical ports. High efficiency therefore requires both
large cooperativity and favorable external coupling. We define the internal
efficiency

\begin{equation}
\eta_{\mathrm{in}} = \frac{\eta}{\eta_e\eta_o},
\label{eq:eta_in}
\end{equation}

which isolates the intrinsic conversion performance of the transducer from
port-coupling losses. In the limit of large cooperativity,
$\eta_{\mathrm{in}} \to 1$, indicating near-ideal internal conversion even
when the total efficiency $\eta$ remains limited by external coupling. This
distinction is particularly useful when comparing platforms whose external
coupling conditions differ substantially.

\subsection{Added Noise}

In practical transduction, thermal fluctuations from microwave waveguides,
internal dissipation, and the intermediate bosonic mode degrade the ideal
pure-loss model~\cite{Gardiner2004,Walls2008}. The input thermal noise
operators satisfy $\delta$-correlated bosonic statistics with mean occupation
$N_{\mu,\mathrm{th}} = (e^{\hbar\omega_\mu/k_BT_\mu}-1)^{-1}$ for mode $\mu$.
The input-referred added noise,

\begin{equation}
N_{\mathrm{add},o} = N_{\mathrm{wg}}
+ \left(\frac{1}{\eta_e}-1\right)N_{e,\mathrm{th}}
+ \frac{1}{\eta_eC_{em}}N_{m,\mathrm{th}},
\label{eq:nadd_full}
\end{equation}

shows that waveguide noise is suppressed by cryogenic cooling, microwave port
noise is minimized in the over-coupled regime $\eta_e \to 1$, and
intermediate-mode noise is reduced by large cooperativity $C_{em} \gg 1$.
Since added noise converts the pure-loss channel into a thermal-loss Gaussian
channel~\cite{Caruso2006,Holevo2001}, achieving nonzero coherent information
requires $N_{\mathrm{add}} \lesssim 1$ in addition to $\eta > 1/2$, as
established in Section~3.

\subsection{Transduction Bandwidth}

The transduction bandwidth defines the spectral range over which effective
microwave-optical conversion is maintained. In cavity-mediated systems it is
set by the effective decay rate of the intermediate mode. For the three
platforms considered here:

\begin{align}
\Delta\omega &= \gamma_m(1 + C_{\mathrm{em}} + C_{\mathrm{om}})
    && \text{(optomechanical)}, \label{eq:bw_om}\\
\Delta\omega &\approx \min(\kappa_e,\kappa_o)
    && \text{(electro-optic)}, \label{eq:bw_eo}\\
\Delta\omega &= \gamma_{\mathrm{mag}} + \Gamma_{\mathrm{mw}} + \Gamma_{\mathrm{opt}}
    && \text{(magneto-optic)}, \label{eq:bw_mo}
\end{align}

where $\gamma_m$ and $\gamma_{\mathrm{mag}}$ are the intrinsic mechanical and
magnon decay rates, respectively. In the optomechanical and magneto-optic cases,
bandwidth grows with cooperativity through dynamical backaction or cooperative
broadening. In the electro-optic case, the absence of a slow mechanical
intermediary means bandwidth is instead limited directly by the cavity
linewidths, which is why electro-optic platforms typically achieve the widest
bandwidths among the three.

\subsection{Operating Temperature}

Operating temperature is a fundamental performance metric because GHz-frequency
modes have energy scales comparable to $k_BT$ even at a few kelvin. The thermal
occupation number

\begin{equation}
n_{\mathrm{th}} = \frac{1}{e^{\hbar\omega/k_BT}-1}
\end{equation}

must satisfy $n_{\mathrm{th}} < 1$ for coherent quantum operation, which
requires operating temperatures of 10--100~mK for mechanical or magnonic
mediators in dilution refrigerators. The platform-specific implications differ:

In \textit{optomechanical} transduction, thermal phonons are the dominant noise
source, with input-referred added noise scaling as $N_{\mathrm{add}} \propto
n_{\mathrm{th}}/C$. Reducing temperature simultaneously lowers noise and
relaxes the cooperativity requirement, making cryogenic operation and strong
coupling jointly essential~\cite{Gardiner2004,Walls2008}.

In \textit{magneto-optic} transduction, thermal magnons must be suppressed to
prevent decoherence during conversion. Cryogenic cooling is therefore equally
essential, and the magnon decay rate $\kappa_m/2\pi$ becomes a critical
parameter since it governs both the thermal occupation and the bandwidth, as
shown in Eq.~(\ref{eq:bw_mo}).

In \textit{electro-optic} transduction, the $\chi^{(2)}$ nonlinear interaction
is not inherently temperature-sensitive, since it involves no low-frequency
mechanical intermediary. Nevertheless, high-performance implementations rely on
superconducting microwave circuitry and therefore still operate at millikelvin
temperatures in practice. This gives electro-optic platforms a potential
advantage for future room-temperature operation, provided that the microwave
circuit requirements can be relaxed through improved materials or all-dielectric
cavity designs~\cite{Khanna2026}.
\section{Transduction Platforms}
\begin{figure}[h!]
    \centering
    \includegraphics[width=0.9\linewidth]{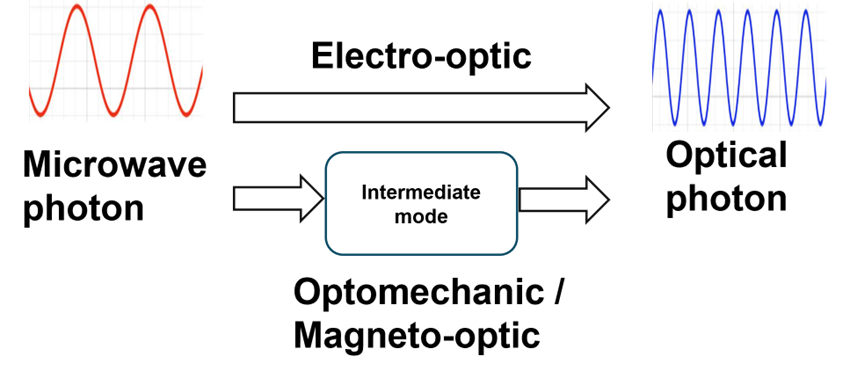}
    \caption{Schematic comparison of the three major quantum transduction
    approaches. Top: zero-stage electro-optic (Pockels) transduction with direct
    photon-photon interaction. Bottom: one-stage transduction using an
    intermediate bosonic mode (phonon or magnon).}
    \label{fig:3approaches}
\end{figure}
In this section we discuss quantum transduction via the zero-stage electro-optic
approach and the one-stage optomechanical and magneto-optic approaches
(demonstrated in Fig.~\ref{fig:3approaches}).

\subsection{Optomechanical Transduction}

\begin{figure*}[h!]
    \centering
    \begin{subfigure}{0.48\textwidth}
        \centering
        \includegraphics[width=\linewidth]{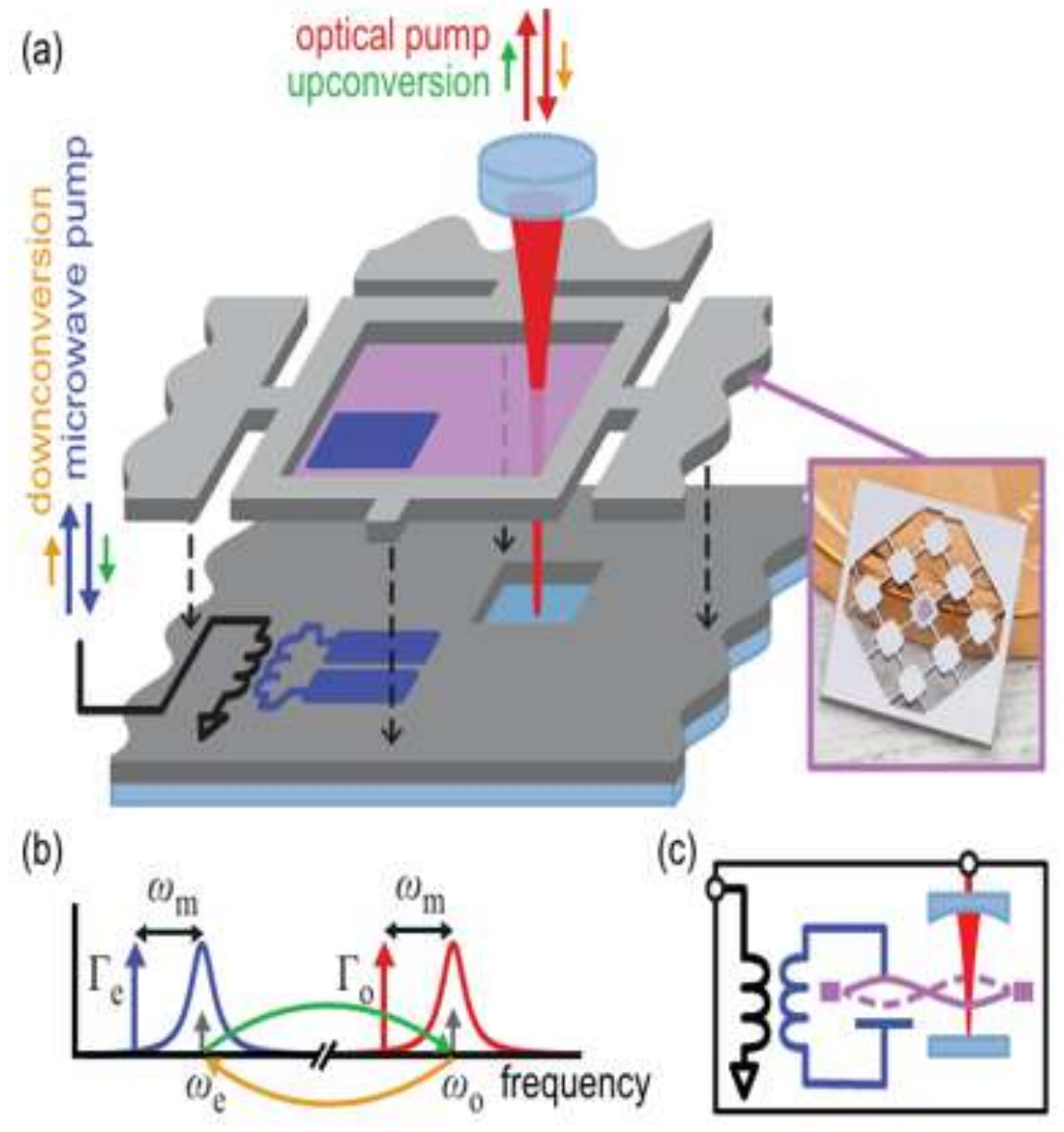}
        \caption{SiN membrane-based platform: a mechanical mode couples a
        microwave LC circuit to an optical Fabry-P\'erot cavity.
        Adapted from~\cite{1_Andrews2014}.}
        \label{fig:membrane}
    \end{subfigure}
    \hfill
    \begin{subfigure}{0.48\textwidth}
        \centering
        \includegraphics[width=\linewidth]{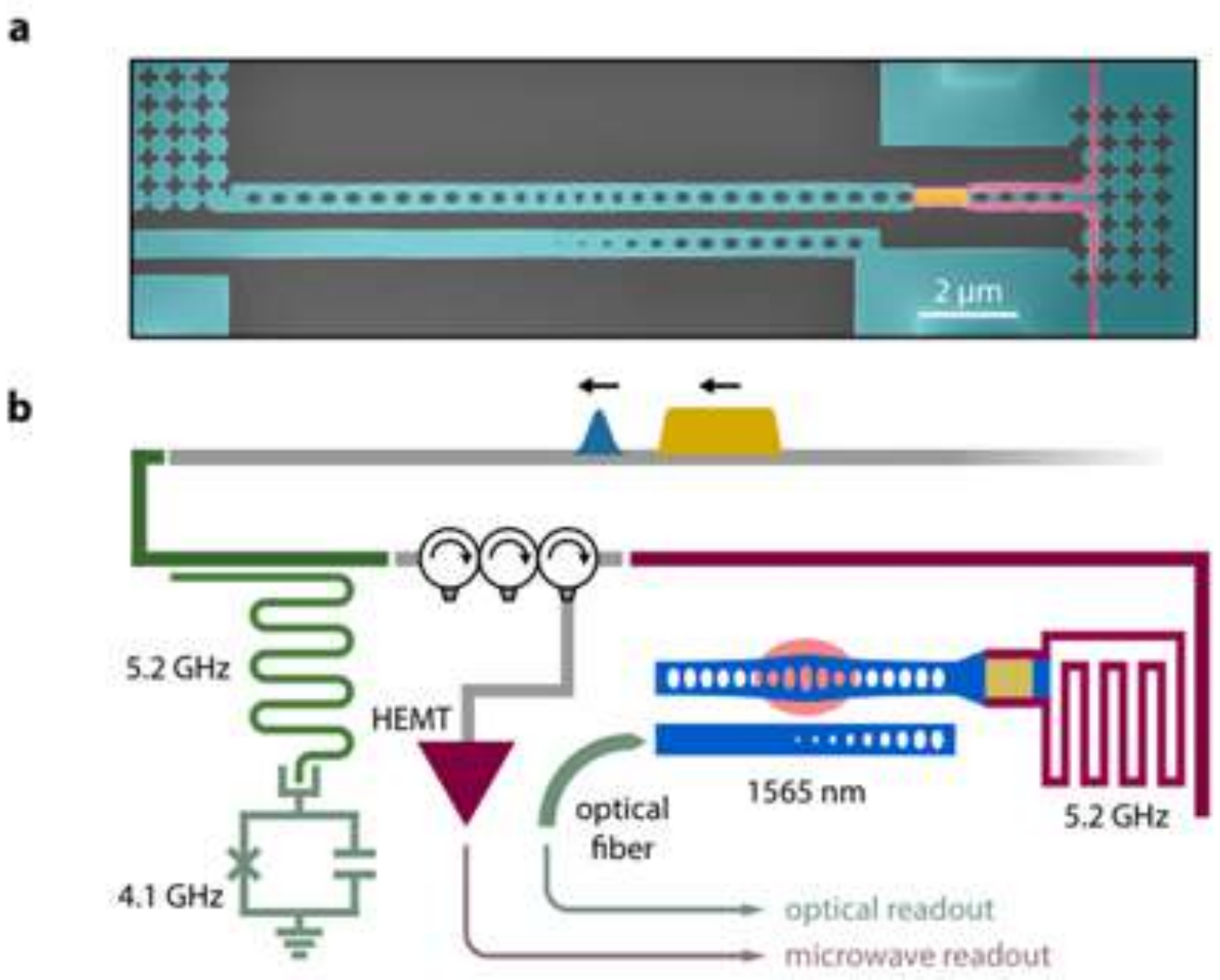}
        \caption{Piezoelectric electro-optomechanical platform: a suspended
        nanobeam OMC couples a 4--5~GHz LC circuit to a 1555~nm optical
        cavity. Adapted from~\cite{2_Vainsencher2016}.}
        \label{fig:piezo}
    \end{subfigure}
    \caption{Optomechanical transduction platforms.}
    \label{fig:optomechanica_membrane_piezo}
\end{figure*}

\begin{figure*}[h!]
    \centering
    \includegraphics[width=\linewidth, trim=0 160 0 160, clip]{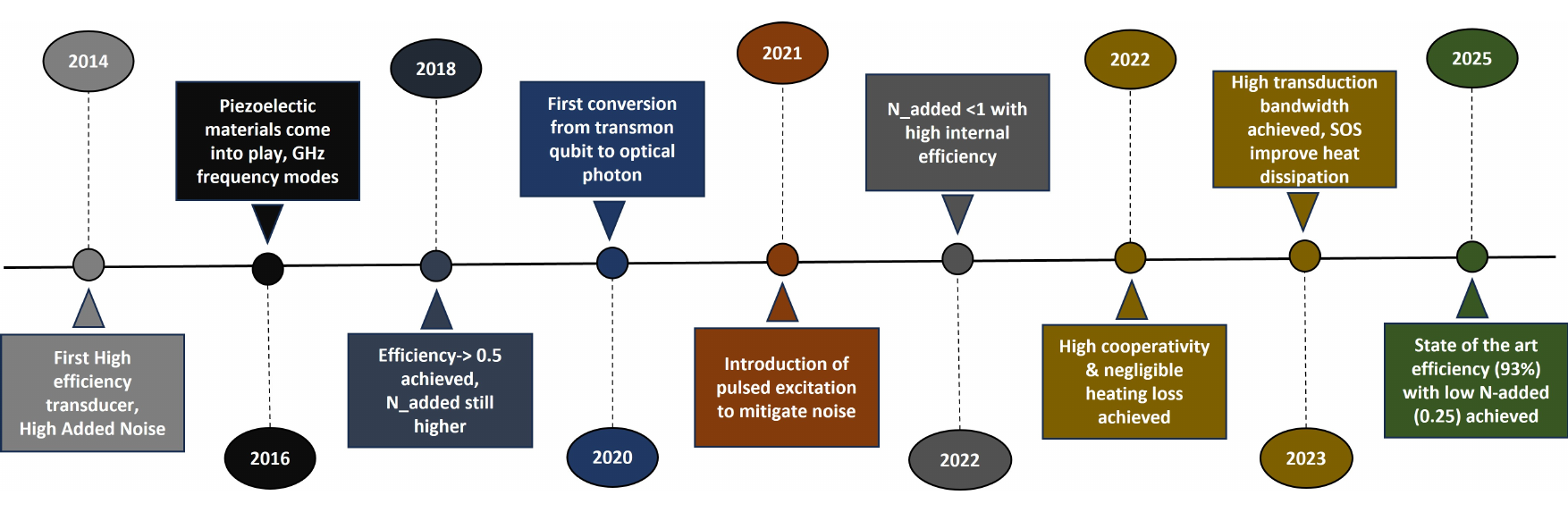}
    \caption{Timeline of key advances in optomechanical microwave-optical quantum
    transduction (2014--2025).}
    \label{fig:optomechanical_timeline}
\end{figure*}

The optomechanical effect couples mechanical motion to light through radiation
pressure and the photoelastic effect. In a transduction context, an intermediate
mechanical resonator simultaneously couples to a microwave circuit via
piezoelectric or capacitive forces and to an optical cavity via radiation
pressure~\cite{Sekine2025Review}. The tripartite system is governed by the
Hamiltonian

\begin{equation}
\hat{H}_{\mathrm{int}} = \hbar g_{em}\!\left(\hat{c}^\dagger\hat{b}
+ \hat{c}\hat{b}^\dagger\right)
+ \hbar g_0\hat{a}^\dagger\hat{a}\!\left(\hat{b}+\hat{b}^\dagger\right),
\label{eq:H_om}
\end{equation}

where $\hat{c}$, $\hat{b}$, and $\hat{a}$ are the annihilation operators for
the microwave, mechanical, and optical modes; $g_{em}$ is the electromechanical
coupling strength; and $g_0$ is the single-photon optomechanical coupling rate.
The single-photon coupling $g_0$ is typically small, so the optical interaction
is linearized by applying a strong drive laser red-detuned from the optical
cavity resonance by $\omega_m$, giving $\omega_d = \omega_o - \omega_m$. This
reduces the optomechanical interaction to an effective beam-splitter Hamiltonian,

\begin{equation}
\hat{H}_{om} \approx \hbar G\!\left(\hat{a}^\dagger\hat{b}
+ \hat{a}\hat{b}^\dagger\right),
\label{eq:H_bs}
\end{equation}

where $G = g_0\sqrt{n_{\mathrm{cav}}}$ is the pump-enhanced coupling rate and
$n_{\mathrm{cav}}$ is the intracavity photon number. Under this Hamiltonian the
mechanical mode acts as a coherent intermediary, swapping excitations between
the microwave and optical fields at a rate determined by the cooperativities
$C_{em}$ and $C_{om}$.

The total efficiency follows Eq.~(\ref{eq:eta}), and is maximized when both
cooperativities are large and the external coupling efficiencies $\eta_e$,
$\eta_o$ approach unity. A dominant noise source is the thermal occupation of
the mechanical mode. Increasing $C_{om}$ by raising the optical pump power
simultaneously heats the mechanical resonator, creating a fundamental trade-off
between efficiency and quantum-limited noise that is one of the central
engineering challenges of the optomechanical platform.

\begin{table*}[h!]
\centering
\caption{Performance comparison of microwave-optical quantum transducers based
on the optomechanical effect. $\eta$: total photon conversion efficiency;
$C_{\mathrm{em}}$, $C_{\mathrm{om}}$: cooperativities; $N_{\mathrm{add}}$:
input-referred added noise (photons); BW: bandwidth; NR: not reported;
CW: continuous-wave; OMC: optomechanical crystal; SOI: silicon-on-insulator.
$^{*}$Internal efficiency.
$^{\S}$Mechanical linewidth $\gamma_m/2\pi$; effective transduction bandwidth
under cooperativity broadening exceeds this value.
$^{\dagger}$Sahu et al.\ (2021) employs an electro-optic (Pockels) mechanism
rather than an optomechanical intermediary; included here to illustrate
the introduction of pulsed excitation as a noise-mitigation strategy.}
\label{tab:optomech_summary}
\scriptsize
\setlength{\tabcolsep}{2.8pt}
\resizebox{\textwidth}{!}{%
\begin{tabular}{l l c c c c c c c c}
\toprule
Reference
& Platform
& $\omega_m$
& $\eta$
& $N_{\mathrm{add}}$
& $C_{\mathrm{em}}$
& $C_{\mathrm{om}}$
& BW
& Temp
& Pump \\
\midrule
Andrews et al.\ (2014)~\cite{1_Andrews2014}
& SiN membrane
& 560 kHz
& $8.6\times10^{-2}$
& ${\sim}10^{3}$
& NR
& NR
& 30 kHz
& 4 K
& CW \\
Vainsencher et al.\ (2016)~\cite{2_Vainsencher2016}
& AlN
& 3.78 GHz
& ${\sim}10^{-8}$
& NR
& --
& $3.3\times10^{-2}$
& ${\sim}1$ MHz
& 300 K
& CW \\
Higginbotham et al.\ (2018)~\cite{3_Higginbotham2018}
& SiN
& 1.47 MHz
& 0.47
& 38
& 66
& 66
& 12 kHz
& $<100\,\mathrm{mK}$
& CW \\
Stockill et al.\ (2022)~\cite{4_Stockill2019}
& GaP
& 2.8 GHz
& 0.02
& 0.55
& --
& $\gg1$
& 191 kHz
& 20 mK
& CW \\
Mirhosseini et al.\ (2020)~\cite{5_Mirhosseini2020}
& SOI
& 5.16 GHz
& $8.8\times10^{-6}$
& 0.57
& --
& NR
& 446 kHz$^{\S}$
& 15 mK
& CW \\
Sahu et al.\ (2021)~\cite{6_Sahu2021}$^{\dagger}$
& LiNbO$_3$
& 8.98 GHz
& 0.15
& 0.16
& 1.2
& NR
& 24 MHz
& 60 mK
& Pulsed \\
Delaney et al.\ (2022)~\cite{7_Delaney2022}
& SiN
& 1.45 MHz
& 0.47
& 3.2
& 1195
& 752
& 220 kHz
& 40 mK
& CW \\
Bl\'esin et al.\ (2024)~\cite{8_Blesin2024}
& SiN/HBARs
& 3.48 GHz
& $1.6\times10^{-5}$
& NR
& --
& NR
& 25 MHz
& 300 K
& CW \\
Zhao et al.\ (2025)~\cite{9_Zhao2025}
& SOI
& 5.07 GHz
& 0.02--0.20
& 0.58--0.94
& NR
& NR
& 88.9 kHz
& mK
& CW \\
Sonar et al.\ (2025)~\cite{10_Sonar2025}
& SOI, 2D OMC
& 10.3 GHz
& $0.93^{*}$
& 0.25
& --
& 1650
& 6 MHz
& 20 mK
& CW+Pulsed \\
\bottomrule
\end{tabular}
}
\end{table*}
\subsubsection{Device Architectures}

Two broad device families have been pursued, illustrated in
Fig.~\ref{fig:optomechanica_membrane_piezo}. The first uses a suspended
dielectric membrane or slab in which the mechanical mode couples capacitively
to a nearby LC circuit and optically to a Fabry-P\'erot or whispering-gallery
cavity~\cite{1_Andrews2014}. The second family, motivated by the need for
stronger electromechanical coupling, uses piezoelectric materials such as AlN
or GaP in which an applied electric field directly produces mechanical strain,
enabling operation at GHz mechanical frequencies without the fabrication
complexity of capacitive gap engineering~\cite{2_Vainsencher2016,4_Stockill2019}.

Optomechanical crystals (OMCs) represent a further advance in this direction.
By patterning periodic holes into a silicon or silicon nitride nanobeam, OMCs
confine both optical and mechanical modes to a sub-micron volume, achieving
single-photon coupling rates $g_0/2\pi$ of several hundred kHz and mechanical
frequencies in the GHz range~\cite{5_Mirhosseini2020,10_Sonar2025}. GHz
mechanical modes are easier to cool to their quantum ground state, since the
sideband-resolved condition $\kappa < \omega_m$ is satisfied, suppressing
unwanted upconversion of thermal phonons into optical
sidebands~\cite{3_Higginbotham2018}. Two-dimensional OMC geometries offer an
additional advantage: the optical and acoustic modes can be mechanically
decoupled by suspending the acoustic resonator away from the optical waveguide,
reducing optical-absorption-induced heating by approximately a factor of
six~\cite{10_Sonar2025}.

\subsubsection{Experimental Progress}

Table~\ref{tab:optomech_summary} and the timeline in
Fig.~\ref{fig:optomechanical_timeline} summarize a decade of experimental
development. Andrews \textit{et al.}~\cite{1_Andrews2014} reported the first
bidirectional microwave-optical transduction using a SiN membrane, achieving
8.6\% total efficiency at 4~K. The low mechanical frequency (560~kHz) produced
large thermal phonon occupation even at cryogenic temperatures, placing the
experiment far from the quantum regime, but the result demonstrated that
mechanical motion is a viable coherent intermediary between microwave and optical
photons.

The introduction of piezoelectric coupling addressed the thermal noise problem
by enabling GHz-frequency mechanical modes. Vainsencher
\textit{et al.}~\cite{2_Vainsencher2016} demonstrated piezo-optomechanical
transducers on silicon-on-insulator, while subsequent GaP-based
devices~\cite{4_Stockill2019} showed that compound semiconductor platforms can
achieve sub-quantum added noise ($N_{\mathrm{add}} = 0.55$) at 20~mK, even at
modest external efficiencies. A key efficiency milestone was reached by
Higginbotham \textit{et al.}~\cite{3_Higginbotham2018}, who implemented a
feed-forward protocol to achieve 47\% total efficiency, approaching the
$\eta = 1/2$ threshold required for positive quantum capacity. The added noise
of 38 photons remained far above the quantum limit, reflecting the fundamental
tension between high cooperativity and low thermal occupation.

The landmark qubit-to-photon transduction experiment of Mirhosseini
\textit{et al.}~\cite{5_Mirhosseini2020} demonstrated that the quantum state
of a superconducting transmon qubit can survive the transduction process.
Using a silicon OMC coupled through an AlN piezoelectric layer, they observed
optical photons correlated with quantum Rabi oscillations of the qubit, with
added noise of 0.57 photons at 15~mK and approximately 1\% total efficiency.
Although the efficiency was low, this experiment provided the first direct
evidence of qubit-state transduction and positioned the SOI OMC as the
leading platform architecture.

Subsequent work addressed the efficiency-noise trade-off through thermal
engineering. Pulsed excitation schemes reduce the average intracavity photon
number and mechanical heating. Bulk acoustic resonator (HBAR) implementations
demonstrated improved heat dissipation at room
temperature~\cite{8_Blesin2024}. Delaney \textit{et al.}~\cite{7_Delaney2022}
achieved 3.2 added noise photons at 47\% efficiency through improved thermal
anchoring and low-loss materials, demonstrating that the two constraints
can be simultaneously approached through careful thermal design rather than
simply increasing pump power.

The current state of the art was established by Sonar
\textit{et al.}~\cite{10_Sonar2025}, who demonstrated 93.1\% internal
phonon-to-photon efficiency with only 0.25 quanta of added noise at 20~mK.
Their two-dimensional silicon OMC featured a mechanically detached acoustic
structure that reduced optical-absorption-induced heating by approximately a
factor of six, suppressing the dominant noise mechanism without sacrificing
coupling strength. This result currently represents one of the closest experimental approaches to quantum-limited optomechanical transduction.

\subsubsection{Outstanding Challenges}

Despite this progress, several challenges remain. Bandwidth is limited by the
mechanical linewidth, typically restricting operation to the kHz--MHz range and
creating compatibility challenges with multiplexed superconducting qubit
architectures that operate in the tens-of-MHz regime~\cite{Lambert2020,Han2021}.
Thin-film LiNbO$_3$ platforms have demonstrated MHz-scale
bandwidths~\cite{6_Sahu2021}, and silicon-on-insulator systems exploiting GHz
mechanical frequencies improve thermal robustness~\cite{9_Zhao2025}, but
simultaneously achieving high efficiency, sub-quantum noise, and wide bandwidth
in a single device remains an open problem~\cite{10_Sonar2025}. Scalable
integration of optomechanical transducers with superconducting qubit chips,
including management of electromagnetic crosstalk and thermal load at millikelvin
temperatures, is a further engineering challenge that must be addressed before
practical deployment in modular quantum computing
architectures~\cite{Ref7_Awschalom2021,8_Blesin2024}.
\subsection{Electro-Optic Transduction}

\begin{figure*}[h!]
    \centering
    \includegraphics[width=\linewidth]{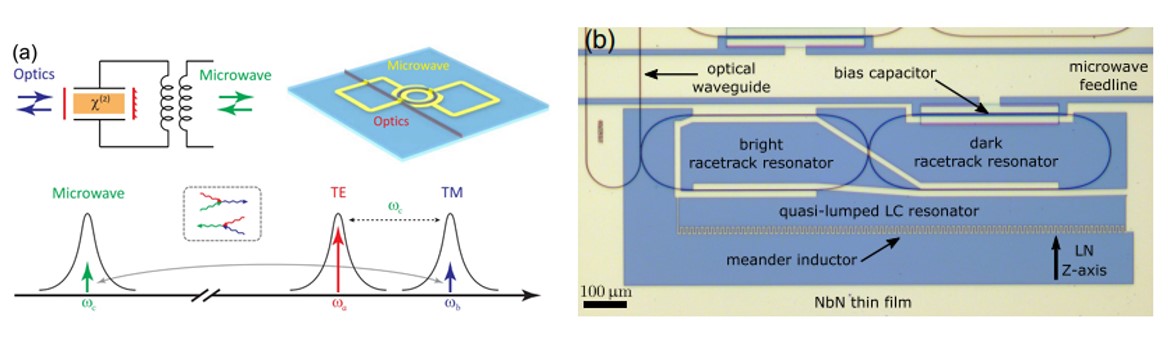}
    \caption{Electro-optic transducers. \textbf{(a)} AlN cavity electro-optic
    transducer setup. Adapted from~\cite{Fan:2022}. \textbf{(b)} Thin-film
    LiNbO$_3$ electro-optic transducer showing racetrack resonators, meander
    microwave inductor, and bias capacitor. Adapted from~\cite{Holzgrafe:20}.}
    \label{fig:eo-transducers}
\end{figure*}
\begin{figure*}[h!]
    \centering
    \includegraphics[width=\textwidth]{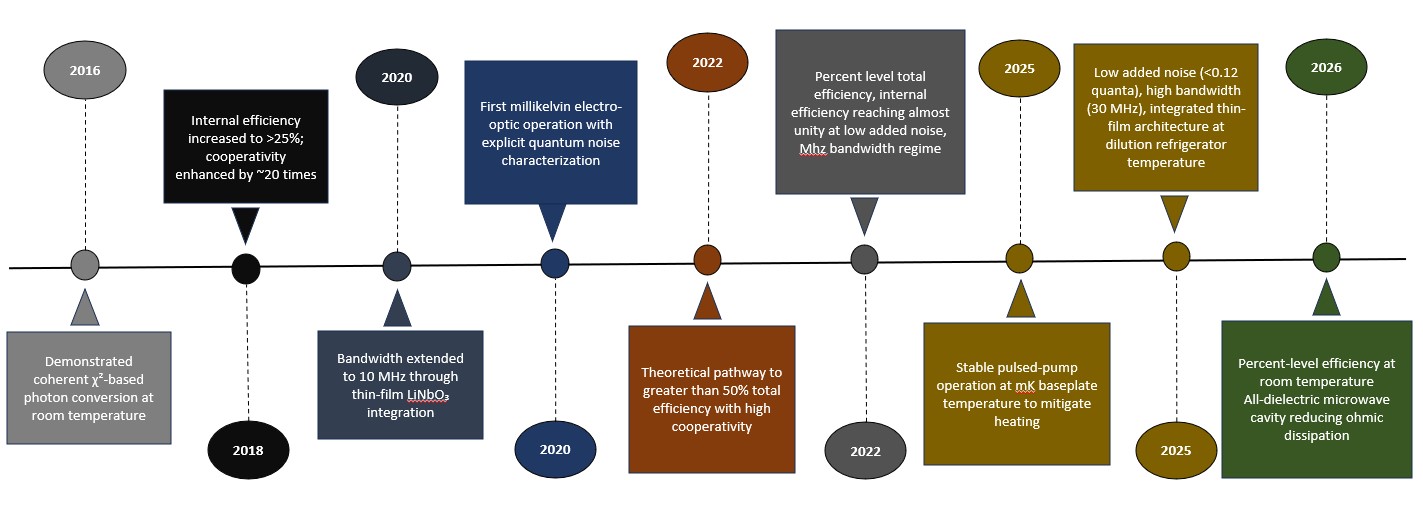}
    \caption{Timeline of advances in electro-optic microwave-optical quantum
    transduction (2016--2026).}
    \label{fig:timeline}
\end{figure*}
\begin{table*}[h!]
\centering
\caption{Performance comparison of microwave-optical quantum transducers based
on the electro-optic (Pockels) effect. $\eta$: total efficiency;
$C_{\mathrm{eo}}$: cooperativity; $N_{\mathrm{add}}$: added noise (photons);
NR: not reported; CW: continuous-wave; WGM: whispering-gallery mode.
$^\dagger$Theoretical projection.}
\label{tab:eo-comparison}
\resizebox{\textwidth}{!}{%
\begin{tabular}{lllcccccl}
\toprule
Reference & Platform & $\eta$ & Internal eff. & $C_{\mathrm{eo}}$ & $N_{\mathrm{add}}$ & Bandwidth & Temperature & Pump \\
\midrule
Khanna et al.\ (2026)~\cite{Khanna2026}  & LiNbO$_3$ bulk cavity & $\sim10^{-2}$ & NR & $(1.7\pm0.8)\times10^{-2}$ & NR & NR & 300 K & CW \\
Warner et al.\ (2025)~\cite{Warner2025} & Thin-film LiNbO$_3$ & 1.18\% & NR & NR & $<0.12$ & 30 MHz & mK & CW \\
Arnold et al.\ (2025)~\cite{Arnold2025} & LiNbO$_3$ WGM & $3\times10^{-3}$ & NR & NR & NR & 10 MHz & 75 mK & Pulsed \\
Wang et al.\ (2022)$^\dagger$~\cite{Wang2022} & Bulk LiNbO$_3$ WGM & 50\% & 93\% & 0.58 & NR & 100 kHz & mK & CW \\
Sahu et al.\ (2022)~\cite{Sahu2022} & Bulk LiNbO$_3$ & 8.7\%, 15\% & 99.5\% & 0.38 & $\sim0.16$ & 18--24 MHz & 60 mK & Pulsed \\
McKenna et al.\ (2020)~\cite{McKenna2020} & Thin-film LiNbO$_3$ & $6.6\times10^{-6}$ & NR & NR & NR & 20 MHz & 1 K & CW \\
Hease et al.\ (2020)~\cite{Hease2020} & Bulk LiNbO$_3$ & $3\times10^{-4}$ & NR & NR & 1.1 & NR & mK & CW \\
Holzgrafe et al.\ (2020)~\cite{Holzgrafe:20} & Thin-film LiNbO$_3$ & $2.7\times10^{-5}$ & NR & NR & NR & 13 MHz & 1 K & CW+Pulsed \\
Fan et al.\ (2018)~\cite{Fan:2022} & AlN cavities & $2.05\pm0.04\%$ & $25.9\pm0.3\%$ & $\sim0.1$ & High & 0.59 MHz & 2 K & CW \\
Rueda et al.\ (2016)~\cite{Rueda2016} & Bulk LiNbO$_3$ WGM & $1.1\times10^{-3}$ & NR & NR & NR & $>1$ MHz & 300 K & CW \\
\bottomrule
\end{tabular}
}
\label{tab:eo-full}
\end{table*}

Quantum transduction via the linear electro-optic (Pockels) effect exploits the
second-order susceptibility $\chi^{(2)}$ in non-centrosymmetric materials to
mediate a direct interaction between microwave and optical photons without an
intermediate mechanical mode. In such materials, the electric polarization
includes a quadratic term

\begin{equation}
P_i^{(2)} = \varepsilon_0\sum_{jk}\chi^{(2)}_{ijk}E_jE_k,
\label{eq:pockels}
\end{equation}

where $\chi^{(2)}_{ijk} \propto r_{ijk}$ is the electro-optic tensor and
$r_{ijk}$ is the Pockels coefficient. When a microwave field and an optical
pump co-propagate in such a medium, this quadratic polarization facilitates
coherent exchange between microwave and optical photons. The key advantage over
optomechanical and magneto-optic platforms is that no slow intermediate
bosonic mode is involved, avoiding the thermal noise and bandwidth limitations
associated with mechanical or magnonic mediators.

In a triply resonant configuration involving one microwave mode $\hat{a}_e$,
one optical pump mode $\hat{a}_p$, and one optical signal mode $\hat{a}_o$,
the interaction Hamiltonian is

\begin{equation}
H_{\mathrm{EO}} = \hbar g_{\mathrm{EO}}
\!\left(\hat{a}_e\hat{a}_p\hat{a}_o^\dagger
+ \hat{a}_e^\dagger\hat{a}_p^\dagger\hat{a}_o\right),
\label{eq:H_EO}
\end{equation}

where the single-photon coupling strength $g_{\mathrm{EO}}$ is determined by
the overlap integral of the microwave and optical mode profiles with
$\chi^{(2)}$. When the pump is treated as a classical coherent field with
amplitude $\sqrt{n_p}$, the substitution $\hat{a}_p \to \sqrt{n_p}$ linearizes
Eq.~(\ref{eq:H_EO}) to the beam-splitter form

\begin{equation}
H_{\mathrm{EO}} = \hbar G_{\mathrm{EO}}
\!\left(\hat{a}_e\hat{a}_o^\dagger + \hat{a}_e^\dagger\hat{a}_o\right),
\label{eq:H_EO_lin}
\end{equation}

where $G_{\mathrm{EO}} = g_{\mathrm{EO}}\sqrt{n_p}$ is the pump-enhanced
coupling rate. This beam-splitter Hamiltonian is formally identical to that of
the optomechanical platform, Eq.~(\ref{eq:H_bs}), but the intermediary is a
virtual photon rather than a phonon, eliminating the associated thermal noise
source. The electro-optic cooperativity is $C_{\mathrm{eo}} =
4G_{\mathrm{EO}}^2/(\kappa_e\kappa_o)$, and the total efficiency follows
Eq.~(\ref{eq:eta}) with $C_{em} = C_{om} = C_{\mathrm{eo}}$.

\subsubsection{Device Architectures}

Practical electro-optic transducers co-locate an optical waveguide or resonator
with a microwave resonator within the nonlinear medium, with the microwave
electrode designed to maximize the radio-frequency electric field overlap with
the optical mode~\cite{Fan:2022,Holzgrafe:20}. Two main resonator geometries
have been pursued, as illustrated in Fig.~\ref{fig:eo-transducers}.

Whispering-gallery-mode (WGM) resonators, typically fabricated from bulk
LiNbO$_3$ crystals, confine optical modes by total internal reflection around
the rim of a disk or sphere. Their high optical quality factors ($Q > 10^8$)
produce strong optical confinement~\cite{Rueda2016}, but the large mode volume
limits the microwave-optical field overlap and therefore the achievable
$g_{\mathrm{EO}}$. Triply resonant conditions in WGM devices are established
by exploiting two optical modes (TE and TM polarizations) whose frequency
spacing matches the microwave resonance frequency, or by engineering the free
spectral range to align a microwave-induced sideband with a cavity
resonance~\cite{Rueda2016,Wang2022}.

Integrated thin-film platforms, most prominently thin-film LiNbO$_3$ on
insulator (LNOI), address the mode-volume limitation by confining both the
optical and microwave modes to sub-micron waveguide cross-sections. The stronger
modal confinement increases $g_{\mathrm{EO}}$ by several orders of magnitude
relative to bulk devices~\cite{McKenna2020,Holzgrafe:20}, and the planar
fabrication process is compatible with superconducting microwave circuit
technology. The microwave resonator is typically implemented as a lumped-element
LC circuit or quasi-lumped meander inductor patterned directly on the chip
alongside the optical racetrack resonators, as shown in
Fig.~\ref{fig:eo-transducers}(b)~\cite{Holzgrafe:20}. AlN thin-film platforms
offer an alternative material choice with lower microwave dielectric loss, at
the cost of a smaller electro-optic coefficient compared to
LiNbO$_3$~\cite{Fan:2022}.

The coherent nature of the electro-optic interaction is empirically verified
through electromagnetically induced transparency (EIT) windows or avoided
crossings in the optical reflection spectrum as the pump frequency or microwave
detuning is varied~\cite{Fan:2022,Sahu2022}. Both signatures confirm that the
microwave and optical modes are hybridized by the $\chi^{(2)}$ interaction
rather than merely co-located on the chip.

\subsubsection{Experimental Progress}

Table~\ref{tab:eo-comparison} and Fig.~\ref{fig:timeline} summarize the
development of electro-optic transducers from room-temperature
proof-of-principle experiments to millikelvin quantum-enabled devices.

Rueda \textit{et al.}~\cite{Rueda2016} established the first coherent,
sideband-resolved microwave-to-optical conversion in a bulk LiNbO$_3$ WGM
resonator at room temperature, achieving $\eta \approx 10^{-3}$ with a
bandwidth exceeding 1~MHz. Although efficiency was low, the experiment
demonstrated that the electro-optic platform can operate in the resolved-sideband
regime and is fundamentally compatible with quantum transduction. Fan
\textit{et al.}~\cite{Fan:2022} subsequently demonstrated the first
superconducting-compatible electro-optic transducer by integrating AlN optical
cavities with superconducting microwave resonators on a single chip at 2~K,
achieving an internal efficiency of 25.9\% and a total efficiency of 2.05\%
with a bandwidth of 0.59~MHz. The observation of an EIT window confirmed
coherent microwave-optical hybridization in a superconducting environment.

The transition to thin-film LiNbO$_3$ platforms enabled a step change in
bandwidth. McKenna \textit{et al.}~\cite{McKenna2020} and Holzgrafe
\textit{et al.}~\cite{Holzgrafe:20} demonstrated cryogenic LNOI devices at
approximately 1~K with bandwidths of 20~MHz and 13~MHz, respectively, though
total efficiencies remained below $10^{-4}$ due to limited cooperativity and
microwave loss. Operations in the millikelvin regime were achieved by Hease
\textit{et al.}~\cite{Hease2020}, who demonstrated bidirectional electro-optic
conversion at the quantum ground state, with added noise of 1.1 photons,
representing the first electro-optic device to approach the quantum-enabled
noise threshold.

The most significant advance in electro-optic transduction came with Sahu
\textit{et al.}~\cite{Sahu2022}, who demonstrated quantum-enabled operation in
a bulk LiNbO$_3$ system at 60~mK. By using short, high-power optical pump
pulses on nanosecond timescales, they achieved near-unity electro-optic
cooperativity, yielding an internal efficiency of 99.5\% and total efficiencies
of 8.7\% and 15\% with added noise as low as 0.16 photons. This result suggests
that the intrinsic electro-optic interaction can achieve near-perfect internal
conversion, and that the remaining gap to unity total efficiency is an external
coupling engineering problem rather than a fundamental limitation of the
platform.

Recent experiments have extended electro-optic transduction to direct qubit
control and readout. Warner \textit{et al.}~\cite{Warner2025} demonstrated
optically driven Rabi oscillations in a superconducting qubit through an
electro-optic transducer operating at millikelvin temperatures with 1.18\%
efficiency and added noise below 0.12 photons at 30~MHz bandwidth, representing an electro-optic demonstration analogous to the milestone achieved in optomechanical transduction by Mirhosseini et al.~\cite{5_Mirhosseini2020} Arnold
\textit{et al.}~\cite{Arnold2025} demonstrated all-optical single-shot qubit
readout at millikelvin temperatures using pulsed electro-optic conversion,
without any active cryogenic microwave equipment, showing that electro-optic
links can replace conventional cryogenic readout chains. Khanna
\textit{et al.}~\cite{Khanna2026} reported a room-temperature all-dielectric
resonant cavity design that reduces microwave ohmic loss by eliminating
metallic electrodes, demonstrating improved stability and a route toward
operation at elevated temperatures.

\subsubsection{Outstanding Challenges}

The central challenge for electro-optic transducers is increasing the total
efficiency while maintaining low added noise and wide bandwidth. The external
coupling efficiency is currently the dominant limiting factor: even when internal
efficiency approaches unity~\cite{Sahu2022}, poor microwave or optical port
coupling suppresses the total efficiency to the percent level. Improving
$\eta_e$ requires either stronger electro-optic coupling to reduce the pump
power needed for high cooperativity, or lower microwave resonator loss to
increase the external coupling fraction~\cite{Fan:2022,Holzgrafe:20}. On the
optical side, fiber-to-chip coupling losses of several dB directly reduce
$\eta_o$ and must be addressed through improved edge coupling or inverse taper
designs~\cite{McKenna2020,Holzgrafe:20}.

Optical pump-induced heating is a secondary but non-negligible challenge. Even
though the $\chi^{(2)}$ interaction is not inherently dissipative, residual
optical absorption in the nonlinear medium deposits heat at millikelvin
temperatures, elevating the microwave mode occupation and degrading
$N_{\mathrm{add}}$~\cite{Hease2020,Sahu2022}. Pulsed pump schemes partially
address this by reducing the average thermal load~\cite{Sahu2022}, but they
sacrifice duty cycle and complicate integration with continuous quantum
communication protocols. All-dielectric cavity designs that eliminate metallic
microwave structures reduce one source of microwave loss~\cite{Khanna2026}, but
achieving simultaneously high optical $Q$ and low microwave loss in a single
monolithic device remains an open fabrication challenge. Theoretical projections
by Wang \textit{et al.}~\cite{Wang2022} suggest that 50\% total efficiency with
93\% internal efficiency is achievable under optimal WGM conditions, providing a
concrete near-term target for the field.
\subsection{Magneto-Optic Transduction}

\begin{figure*}[h!]
    \centering
    \includegraphics[width=\linewidth]{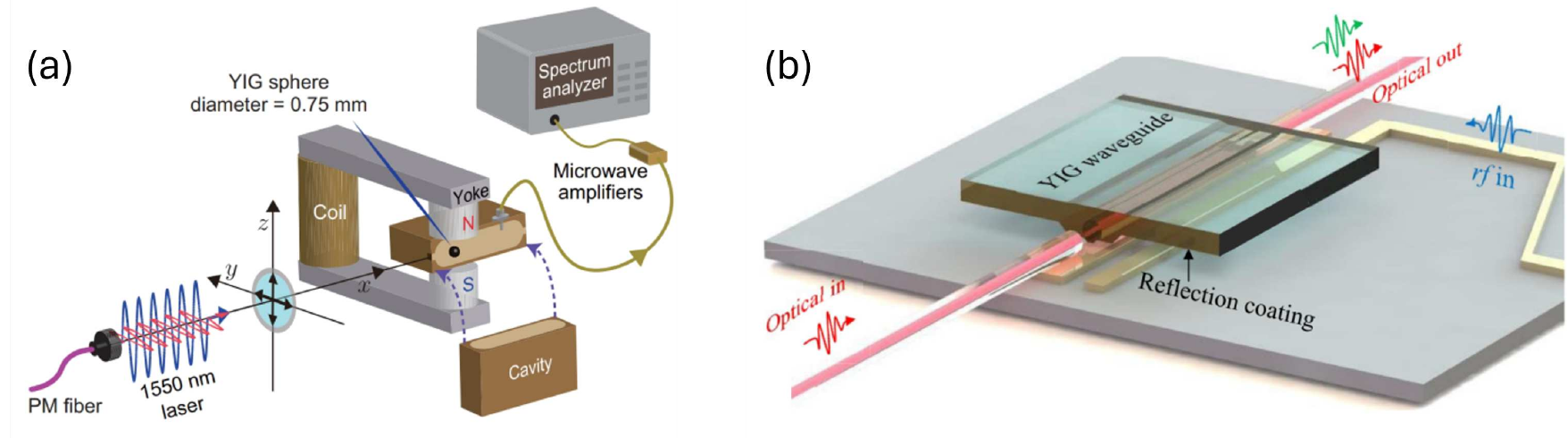}
    \caption{Magneto-optic transduction platforms. \textbf{(a)} YIG sphere in
    a microwave cavity coupled to a 1550~nm laser via the Faraday effect.
    Adapted from~\cite{Hisatomi2016}. \textbf{(b)} Integrated optomagnonic
    waveguide device. Adapted from~\cite{Zhu2020}.}
    \label{fig:magneto-opto}
\end{figure*}

\begin{figure*}[h!]
    \centering
    \includegraphics[width=\linewidth, trim=0 80 0 60, clip]{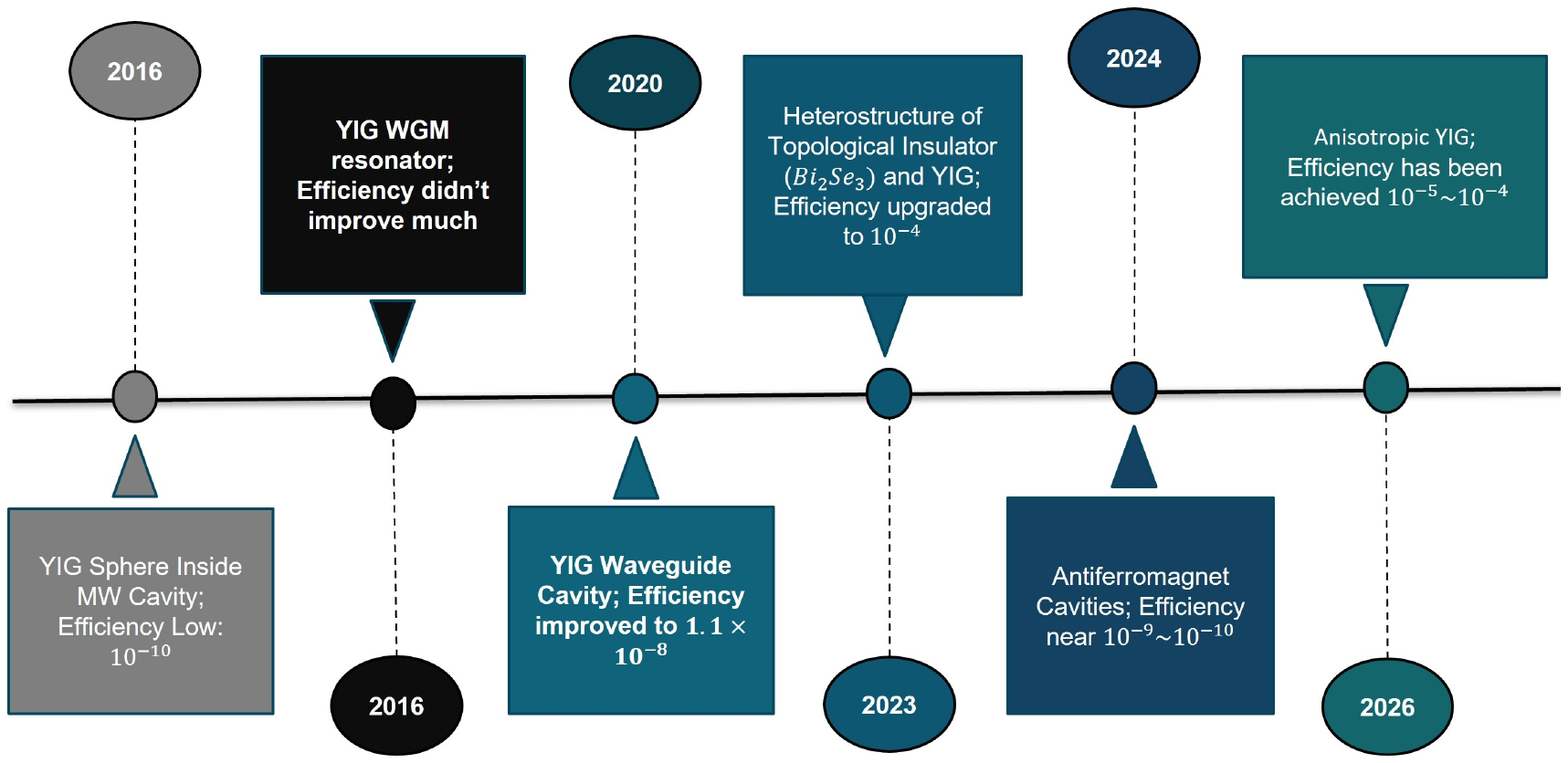}
    \caption{Timeline of advances in magneto-optic microwave-optical quantum
    transduction (2016--2026).}
    \label{fig:time_magneto-opto}
\end{figure*}

\begin{table*}[t]
\centering
\caption{Performance comparison of microwave-optical quantum transducers based
on the magneto-optic effect. $\eta$: photon conversion efficiency;
$C_{\mathrm{em}}$: microwave-magnon cooperativity; $C_{\mathrm{om}}$:
optomagnonic cooperativity; $\kappa_m/2\pi$: intrinsic magnon decay rate;
NR: not reported; RT: room temperature; pred.: predicted/theoretical.}
\label{tab:MOcomparison}

\scriptsize
\setlength{\tabcolsep}{3pt}

\resizebox{\textwidth}{!}{%
\begin{tabular}{l l c c c c c c}
\toprule
Reference & System & $\eta$ & $C_{\mathrm{em}}$ & $C_{\mathrm{om}}$ & Bandwidth & Temp. & $\kappa_m/2\pi$ \\
\midrule

Hisatomi et al.\ (2016)~\cite{Hisatomi2016}
& YIG sphere + MW cavity
& $\sim10^{-10}$
& $\sim10^{2}$--$10^{3}$
& NR
& NR
& $\sim300$ K
& $\sim1$ MHz \\

Osada et al.\ (2016)~\cite{Osada2016}
& YIG WGM resonator
& $7\times10^{-14}$
& NR
& NR
& NR
& $\sim300$ K
& NR \\

Zhang et al.\ (2016)~\cite{Zhang2016}
& YIG cavity
& $1.7\times10^{-15}$
& NR
& $5.4\times10^{-7}$
& NR
& $\sim300$ K
& NR \\

Zhu et al.\ (2020)~\cite{Zhu2020}
& YIG waveguide cavity
& $1.1\times10^{-8}$
& 0.87
& $4.1\times10^{-7}$
& 16.1 MHz
& $\sim300$ K
& 3.25 MHz \\

Sekine et al.\ (2023)~\cite{Sekine2023TI}
& Topological insulator + YIG
& $\sim10^{-4}$ (pred.)
& NR
& Intrinsically enhanced
& THz--GHz
& RT
& NR \\

Sekine et al.\ (2024)~\cite{Sekine2024AFM}
& Antiferromagnet + cavities
& $\sim10^{-9}$
& NR
& NR
& THz regime
& $\sim300$ K
& 100 MHz \\

Xie et al.\ (2026)~\cite{Xie2026}
& Anisotropic YIG (squeezed)
& $10^{-5}$--$10^{-4}$ (pred.)
& NR
& Enhanced via $\cosh r$
& MHz
& RT
& Model dep. \\

\bottomrule
\end{tabular}%
}
\end{table*}

Magneto-optic quantum transduction uses magnons, the collective spin-wave
excitations of a ferromagnetic material, as the intermediary bosonic mode
between microwave and optical photons. The magnon frequency
$\omega_m = \gamma B$ is set by the external static magnetic field $B$ and the
gyromagnetic ratio $\gamma$, providing a continuously tunable resonance
condition that is absent in geometrically fixed mechanical or electro-optic
platforms. This tunability suggests a distinctive advantage of the magneto-optic
approach, since it allows the transducer to be brought into resonance with a
given superconducting qubit frequency without redesigning the device.

The microwave-magnon interaction originates from Zeeman coupling and reduces in
a cavity to the beam-splitter Hamiltonian

\begin{equation}
H_{\mathrm{em}} = \hbar g_{\mathrm{em}}
\!\left(\hat{a}_e^\dagger\hat{a}_m + \hat{a}_e\hat{a}_m^\dagger\right),
\label{eq:H_em}
\end{equation}

where $\hat{a}_m$ is the magnon annihilation operator and $g_{\mathrm{em}}$ is
the microwave-magnon coupling strength. In a ferromagnet with $N_s$ spins,
collective enhancement gives $g_{\mathrm{em}} \propto \sqrt{N_s}$, so the
microwave-magnon cooperativity $C_{\mathrm{em}} \propto N_s$ can readily exceed
unity in macroscopic YIG samples~\cite{Hisatomi2016}. This strong microwave coupling is generally regarded as one of the principal strengths of the magneto-optic platform.

The magnon-optical interaction arises from the Faraday effect, in which the
magnetization of the medium rotates the polarization of a transmitted optical
field. The interaction Hamiltonian is

\begin{equation}
H_{\mathrm{MO}} = -\frac{i\varepsilon_0 f}{4}
\int d^3r\,\mathbf{M}\cdot(\mathbf{E}^*\times\mathbf{E}),
\label{eq:H_Faraday}
\end{equation}

where $f$ is the magneto-optical coupling constant and $\mathbf{M}$ is the
magnetization. Linearizing about a coherent intracavity optical field with mean
photon number $\bar{n}_{\mathrm{cav}}$ yields the effective beam-splitter
Hamiltonian

\begin{equation}
H_{\mathrm{MO}} = \hbar G_{\mathrm{MO}}
\!\left(\hat{a}_o^\dagger\hat{a}_m + \hat{a}_o\hat{a}_m^\dagger\right),
\label{eq:H_MO_lin}
\end{equation}

where $G_{\mathrm{MO}} = g_{\mathrm{MO},0}\sqrt{\bar{n}_{\mathrm{cav}}}$ is
the pump-enhanced optomagnonic coupling rate. The single-photon coupling
strength is

\begin{equation}
g_{\mathrm{MO},0} = \frac{c\,\theta_F}{4\sqrt{2\varepsilon_r N_s}},
\label{eq:g_MO}
\end{equation}

where $c$ is the speed of light in vacuum, $\theta_F$ is the Faraday rotation
angle per unit length, and $\varepsilon_r$ is the relative permittivity of the
medium. The critical consequence of Eq.~(\ref{eq:g_MO}) is that
$g_{\mathrm{MO},0} \propto 1/\sqrt{N_s}$: the same large spin number that
strengthens microwave coupling simultaneously weakens optical coupling. The
optomagnonic cooperativity therefore satisfies $C_{\mathrm{om}} \ll 1$ for all
bulk ferromagnets studied to date, and the total efficiency under triple
resonance,

\begin{equation}
\eta \propto C_{\mathrm{em}}\,C_{\mathrm{om}},
\label{eq:eta_mo}
\end{equation}

is fundamentally suppressed by the weakness of the Faraday interaction. This
persistent asymmetry between $C_{\mathrm{em}} \gg 1$ and $C_{\mathrm{om}} \ll
1$ defines the central bottleneck of the magneto-optic platform, and all
recent theoretical and experimental effort is directed at closing this gap.

\subsubsection{The Role of the Magnon Decay Rate}

A key contribution of this review is the identification of the intrinsic magnon
decay rate $\kappa_m/2\pi$ as a cross-platform parameter that simultaneously
governs efficiency, bandwidth, cooperativity, and thermal noise in magneto-optic
systems, and which has not been consistently reported or compared across prior
experimental studies. The bandwidth under triple resonance is

\begin{equation}
\Delta\omega = \kappa_m(1 + C_{\mathrm{em}} + C_{\mathrm{om}}),
\label{eq:bw_mo_detail}
\end{equation}

showing that bandwidth can be enhanced through cooperative broadening even when
the intrinsic decay is slow. YIG is distinguished by an exceptionally low
intrinsic magnon decay rate $\kappa_m/2\pi \approx 1$~MHz and a tunable
ferromagnetic resonance frequency near 10~GHz, making it the material of choice
for low-noise magneto-optic
transduction~\cite{Tabuchi2018,Zhang2014,rameshti2022cavity}. Newer platforms
such as antiferromagnets exhibit substantially higher decay rates
($\kappa_m/2\pi \approx 100$~MHz), which reduces efficiency through
Eq.~(\ref{eq:eta_mo}) but enables THz-band operation and removes the need for
a static bias field~\cite{Sekine2024AFM}.

\subsubsection{Device Architectures}

Three main device geometries have been explored, each representing a different
strategy for increasing $C_{\mathrm{om}}$.

The simplest architecture places a polished YIG sphere inside a microwave
cavity and couples it to an optical mode via free-space or fiber
optics~\cite{Hisatomi2016,Tabuchi2018}. YIG spheres support high-$Q$ magnon
modes ($Q > 10^4$) and are straightforward to fabricate, but their large mode
volume ($\sim$mm diameter) produces poor optical confinement and a
correspondingly small Faraday interaction~\cite{Hisatomi2016}. WGM resonator
implementations attempt to improve optical confinement by exploiting total
internal reflection within the YIG sphere itself~\cite{Osada2016}, but the
refractive index of YIG ($n \approx 2.2$) limits the achievable optical quality
factor and the field overlap remains insufficient for practical transduction.

Integrated waveguide cavity devices, pioneered by Zhu
\textit{et al.}~\cite{Zhu2020}, address the mode-volume problem by confining
the optical field within a sub-micron YIG waveguide evanescently coupled to a
tapered optical fiber, while the microwave mode is confined by a surrounding
resonant cavity structure. The reduced optical mode volume increases the
per-spin Faraday interaction, improving $C_{\mathrm{om}}$ by two to three
orders of magnitude relative to sphere-based devices~\cite{Zhu2020}, though it
remains far below unity.

Heterostructure platforms represent the most recent architectural direction.
By combining YIG with materials that exhibit stronger or topologically protected
magneto-optical responses, these devices aim to decouple the $N_s$ dependence
of $g_{\mathrm{MO},0}$ from the material volume. Topological insulator
overlayers~\cite{Sekine2023TI}, antiferromagnetic films~\cite{Sekine2024AFM},
and geometrically anisotropic YIG structures~\cite{Xie2026} are all active
areas of investigation, as discussed below.

\subsubsection{Experimental Progress}

Table~\ref{tab:MOcomparison} and the timeline in Fig.~\ref{fig:time_magneto-opto}
summarize the experimental development of magneto-optic transducers. The first
three demonstrations appeared simultaneously in 2016 and collectively
established the key physics of the platform. Hisatomi
\textit{et al.}~\cite{Hisatomi2016} demonstrated bidirectional microwave-optical
conversion via YIG magnons in a sphere-plus-cavity geometry, achieving
$\eta \sim 10^{-10}$ and $C_{\mathrm{em}} \sim 10^2$--$10^3$, confirming that
the microwave interface is not the limiting factor. Osada
\textit{et al.}~\cite{Osada2016} implemented a WGM resonator geometry and
reported $\eta = 7\times10^{-14}$, while Zhang
\textit{et al.}~\cite{Zhang2016} measured $C_{\mathrm{om}} = 5.4\times10^{-7}$
in a YIG cavity system, providing the first direct experimental confirmation
that $C_{\mathrm{om}} \ll 1$ is the bottleneck consistent with
Eq.~(\ref{eq:g_MO}).

The integrated waveguide cavity of Zhu \textit{et al.}~\cite{Zhu2020}
represented a significant experimental advance, improving
efficiency to $1.1\times10^{-8}$ and demonstrating a bandwidth of 16.1~MHz.
The bandwidth exceeds $\kappa_m/2\pi = 3.25$~MHz due to cooperative broadening
consistent with Eq.~(\ref{eq:bw_mo_detail}), confirming that bandwidth is not
fundamentally limited by intrinsic magnon decay and can be engineered through
coupling. Despite the two-to-three order-of-magnitude efficiency improvement,
$C_{\mathrm{om}} = 4.1\times10^{-7}$ remained essentially unchanged from the
2016 sphere results, demonstrating that improved optical confinement alone is
insufficient and that a materials-level enhancement of the Faraday interaction
is required.
\subsubsection{Emerging Theoretical Directions}

Since engineering $C_{\mathrm{om}}$ through device geometry has reached its
practical limits, recent theoretical work targets the Faraday interaction
itself through three distinct strategies.

Magnetic anisotropy in YIG thin films establishes a squeezed vacuum
for the magnonic ground state, exponentially enhancing the effective
optomagnonic coupling by a factor of $\cosh r$, where $r$ is the
squeezing parameter determined by the anisotropy
strength~\cite{Xie2026}. Xie \textit{et al.}\ predict that this
anisotropy-induced magnon squeezing can raise the microwave-to-optical
conversion efficiency by several orders of magnitude, reaching
$10^{-5}$-$10^{-4}$ for large squeezing.

Topological insulator-YIG heterostructures exploit the surface states of
materials such as Bi$_2$Se$_3$, which exhibit a Faraday rotation that is
quantized by the topological surface conductance and therefore independent of
film thickness. Sekine \textit{et al.}~\cite{Sekine2023TI} predict that
coupling such a heterostructure to a YIG magnon mode can intrinsically boost
$C_{\mathrm{om}}$ without increasing the ferromagnet volume, potentially circumventing the
$1/\sqrt{N_s}$ scaling that limits bulk devices.

Antiferromagnetic systems offer a third route by shifting the magnon resonance
into the THz regime through exchange coupling, which is orders of magnitude
stronger than the dipolar coupling that sets the GHz resonance frequency of
ferromagnets. Sekine \textit{et al.}~\cite{Sekine2024AFM} showed that
antiferromagnet-cavity systems can operate without a bias magnetic field,
potentially suppressing thermal magnon occupation and simplifying integration
with superconducting circuits, though current efficiency predictions remain near
$10^{-9}$.

\subsubsection{Outstanding Challenges}

The magneto-optic platform faces two qualitatively distinct challenges. The
first is the $C_{\mathrm{om}}$ problem: raising the optomagnonic cooperativity
from $\sim10^{-7}$~\cite{Zhang2016,Zhu2020} to values approaching unity
requires either a fundamentally stronger Faraday interaction than is available
in bulk YIG, or a device geometry that dramatically reduces the optical mode
volume while maintaining a large spin ensemble. Neither has been demonstrated
experimentally, and the theoretical proposals described above remain to be
validated~\cite{Xie2026,Sekine2023TI,Sekine2024AFM}. The second challenge is
operating temperature: unlike the optomechanical and electro-optic platforms,
where millikelvin operation has been demonstrated~\cite{10_Sonar2025,Sahu2022},
all magneto-optic experiments to date have been conducted at room
temperature~\cite{Hisatomi2016,Osada2016,Zhang2016,Zhu2020}. Cryogenic
operation is expected to suppress thermal magnon occupation and reduce
$N_{\mathrm{add}}$, but the integration of optical fiber coupling and microwave
cavities at millikelvin temperatures in the presence of strong static magnetic
fields presents a significant experimental
complexity~\cite{rameshti2022cavity,Tabuchi2018}.

Despite these limitations, the magneto-optic platform occupies a unique niche.
Its inherent non-reciprocity, arising from the time-reversal symmetry breaking
of the Faraday effect~\cite{Hisatomi2016}, offers functionality unavailable in
optomechanical or electro-optic devices: directional photon routing, optical
isolation, and magnon-based quantum memory are all natural extensions of the
platform~\cite{ZhuCirculator2020,rameshti2022cavity}. If $C_{\mathrm{om}}$ can
be raised by even two orders of magnitude through the emerging theoretical
strategies~\cite{Xie2026,Sekine2023TI}, magneto-optic transducers could become
competitive with the other platforms for non-reciprocal quantum network
applications where directionality is as important as
efficiency~\cite{ZhuCirculator2020}.

\section{Cross-Platform Comparison}

\begin{figure*}[t]
    \centering
    \includegraphics[width=\textwidth]{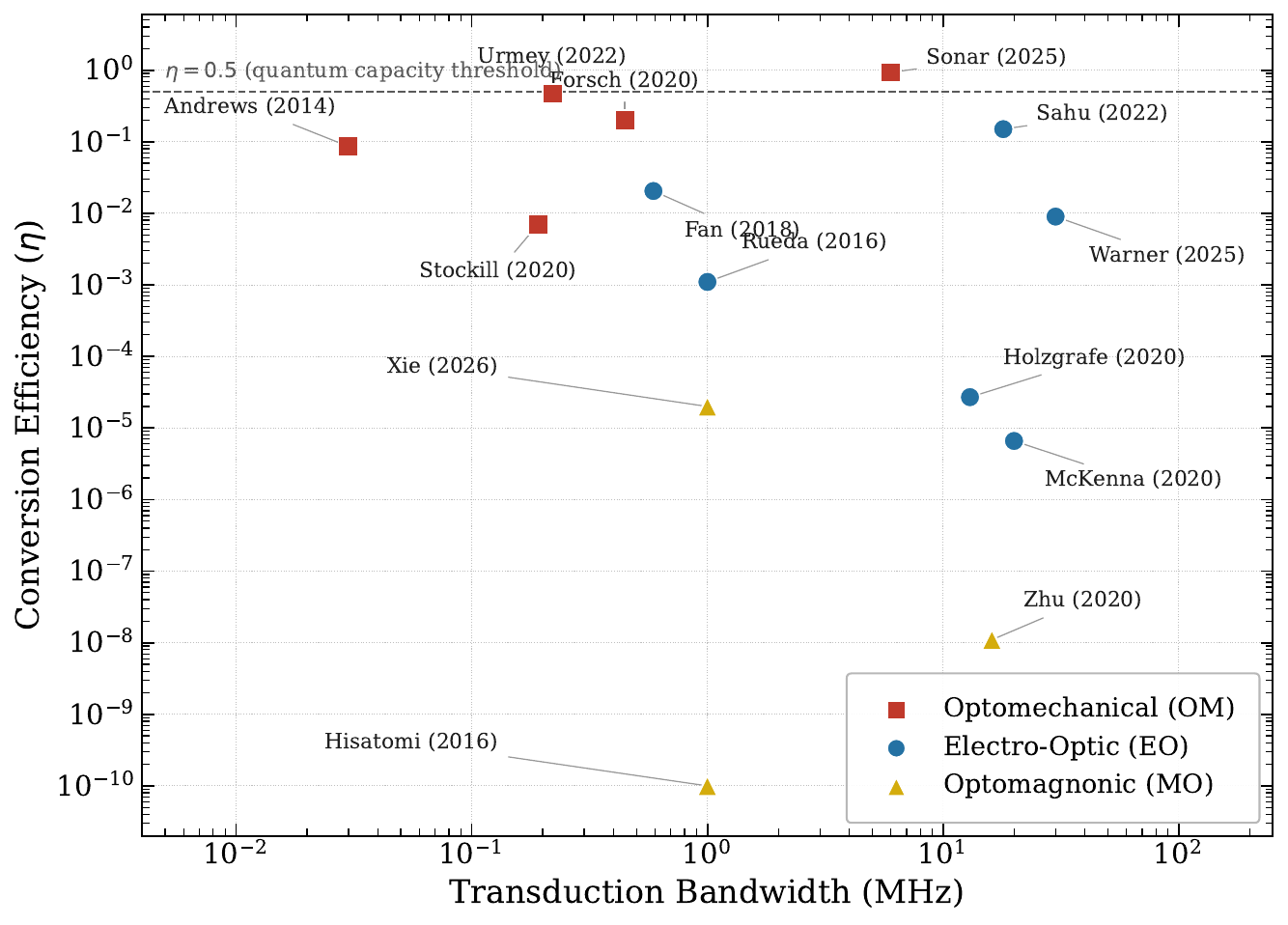}
    \caption{Comparison of microwave-optical quantum transduction platforms in
    the efficiency-bandwidth plane. Optomechanical systems (squares) achieve the
    highest conversion efficiencies; electro-optic systems (circles) offer the
    widest bandwidths; magneto-optic systems (triangles) currently show the
    lowest efficiencies but offer unique non-reciprocal functionality. The dashed
    line marks the $\eta = 0.5$ threshold required for positive quantum channel
    capacity.}
    \label{fig:current_state}
\end{figure*}

Having reviewed each platform individually, we now compare 
their performance directly across the key metrics established 
in Section~4. Fig.~\ref{fig:current_state} plots experimental results in the
efficiency-bandwidth plane, illustrating the complementary operating regimes of
the three platforms. Table~\ref{tab:platform_summary} provides a consolidated
quantitative comparison of the key metrics discussed in this review.

\begin{table}[t]
\centering
\caption{Consolidated cross-platform comparison of microwave-optical quantum
transduction. Values represent the best experimentally demonstrated results
unless otherwise noted. $^\dagger$: theoretical projection;
$^\ddagger$: internal efficiency; NR: not reported.}
\label{tab:platform_summary}

\scriptsize
\setlength{\tabcolsep}{2.5pt}

\resizebox{0.48\textwidth}{!}{%
\begin{tabular}{lccc}
\toprule
Metric & Optomechanical & Electro-Optic & Magneto-Optic \\
\midrule

Best $\eta_{\mathrm{in}}$
& $93\%$~\cite{10_Sonar2025}
& $99.5\%$~\cite{Sahu2022}
& NR \\

Best total $\eta$
& $47\%$~\cite{3_Higginbotham2018,7_Delaney2022}
& $15\%$~\cite{Sahu2022}
& $\sim10^{-8}$~\cite{Zhu2020} \\

Best $N_{\mathrm{add}}$
& $0.25$~\cite{10_Sonar2025}
& $0.12$~\cite{Warner2025}
& NR \\

Bandwidth
& kHz--MHz
& MHz--100 MHz
& MHz--THz \\

Min.\ temperature
& 20 mK~\cite{10_Sonar2025}
& 60 mK~\cite{Sahu2022}
& $\sim$300 K~\cite{Hisatomi2016} \\

Intermediate mode
& Phonon
& None
& Magnon \\

Non-reciprocal
& No
& No
& Yes~\cite{Hisatomi2016,Zhu2020} \\

Field tunable
& No
& No
& Yes~\cite{Hisatomi2016} \\

\bottomrule
\end{tabular}%
}
\end{table}
\textbf{Optomechanical systems} currently lead in conversion efficiency, with
state-of-the-art devices achieving 93\% internal efficiency and sub-quantum
added noise of 0.25 photons at millikelvin temperatures~\cite{10_Sonar2025}.
Their principal limitation is bandwidth, constrained to the kHz--MHz range by
the mechanical linewidth~\cite{Lambert2020,Han2021}, which restricts
compatibility with multiplexed superconducting qubit architectures. They appear particularly well suited to applications where conversion fidelity is the dominant
requirement, such as high-fidelity qubit state readout and entanglement
generation between remote nodes~\cite{5_Mirhosseini2020,7_Delaney2022}.

\textbf{Electro-optic systems} occupy the high-bandwidth region of the
efficiency-bandwidth space, with demonstrated bandwidths of 10--100~MHz and
internal efficiencies approaching unity~\cite{Sahu2022,Wang2022}. The absence
of a mechanical intermediary eliminates the associated thermal noise source and
enables faster response times. Total external efficiencies remain in the percent
range, limited by port-coupling losses rather than intrinsic
conversion~\cite{Sahu2022}, and recent experiments have demonstrated direct
optical control and readout of superconducting
qubits~\cite{Warner2025,Arnold2025}. These platforms may be specially suitable for high-speed multiplexed quantum links and scalable modular
architectures~\cite{Ref7_Awschalom2021}.

\textbf{Magneto-optic systems} currently achieve the lowest efficiencies
($10^{-10}$--$10^{-8}$)~\cite{Hisatomi2016,Zhu2020}, limited fundamentally by
the weak optomagnonic cooperativity $C_{\mathrm{om}} \ll
1$~\cite{Osada2016,Zhang2016}. However, they offer two capabilities unavailable
in the other platforms: inherent non-reciprocity arising from time-reversal
symmetry breaking, and continuous frequency tunability via an external magnetic
field~\cite{Hisatomi2016,rameshti2022cavity}. These properties make
magneto-optic devices natural candidates for non-reciprocal quantum network
components such as circulators and directional
routers~\cite{ZhuCirculator2020}, provided efficiency can be raised through
emerging theoretical strategies~\cite{Xie2026,Sekine2023TI,Sekine2024AFM}.

No single platform simultaneously satisfies the combined requirements of high
efficiency, low added noise, wide bandwidth, and room-temperature operation.
Future quantum network architectures will likely exploit complementary
combinations: optomechanical transducers for high-fidelity local interfaces,
electro-optic links for long-distance high-bandwidth interconnects, and
magneto-optic elements for non-reciprocal routing and
isolation~\cite{Ref6_Lauk2020,Ref7_Awschalom2021,Sekine2025Review}.

\section{Outlook and Future Directions}

The results surveyed in this review identify four interconnected challenges
that must be addressed before microwave-optical transduction can underpin
practical quantum networks.

\textbf{Simultaneous optimization of efficiency, noise, and bandwidth.} No
current platform satisfies all three requirements at once. In optomechanical
systems, the pump-power-induced heating that limits noise performance must be
decoupled from coupling strength, for example through phononic shielding,
pulsed operation, or two-dimensional OMC geometries that thermally isolate the
acoustic mode from the optical absorption region~\cite{10_Sonar2025}. In
electro-optic systems, increasing $g_{\mathrm{EO}}$ through improved mode
overlap and new nonlinear materials with larger Pockels coefficients, such as
BaTiO$_3$ and SrTiO$_3$, could raise total efficiency while relaxing the
pump-power requirement~\cite{Mohl2025,Khanna2026}. In magneto-optic systems,
raising $C_{\mathrm{om}}$ by even two orders of magnitude through
anisotropy-induced magnon squeezing~\cite{Xie2026} or topological
enhancement~\cite{Sekine2023TI}

\textbf{On-chip integration with superconducting qubits.} Practical deployment
requires transducers to be co-integrated with superconducting processors,
minimizing microwave loss, electromagnetic crosstalk, and thermal load at
millikelvin temperatures~\cite{Ref7_Awschalom2021}. Electro-optic platforms are
the most advanced in this direction, with recent experiments demonstrating
optical Rabi oscillation drive and all-optical single-shot readout of
superconducting qubits without any active cryogenic microwave
equipment~\cite{Warner2025,Arnold2025}. Optomechanical integration faces greater
challenges due to the sensitivity of mechanical resonators to vibration and the
thermal management demands of co-located optical and microwave
circuitry~\cite{10_Sonar2025,8_Blesin2024}.

\textbf{Relaxing cryogenic requirements.} All high-performance transducers
currently require millikelvin operation~\cite{10_Sonar2025,Sahu2022,Warner2025}.
Electro-optic transduction is the most promising candidate for relaxing this
constraint, since the $\chi^{(2)}$ interaction is not inherently
temperature-limited. All-dielectric resonant cavity designs that eliminate
metallic microwave structures reduce ohmic heating at elevated
temperatures~\cite{Khanna2026}, and represent an early step toward
room-temperature or near-room-temperature operation. New material platforms with
large Pockels coefficients and cryogenic compatibility, such as
BaTiO$_3$~\cite{Mohl2025}, further broaden the materials space available for
this goal.

\textbf{Non-reciprocal quantum network components.} Quantum circulators,
isolators, and directional amplifiers are essential components of any
large-scale quantum network, and conventional ferrite-based microwave
circulators are bulky, magnetic-field-dependent, and incompatible with
superconducting circuit integration~\cite{ZhuCirculator2020}. The inherent
non-reciprocity of magneto-optic transduction, combined with the emerging
capability to extend operation into the THz regime via antiferromagnetic
systems~\cite{Sekine2024AFM}, motivates continued investment in this platform
beyond its current efficiency limitations. Magnon-based cavity systems have
been proposed as a route to high-isolation, low-insertion-loss microwave
circulators compatible with superconducting
platforms~\cite{ZhuCirculator2020,rameshti2022cavity}, and realizing the
optical analogue through magneto-optic transduction remains an important open
direction.
\section{Conclusion}

Microwave-to-optical quantum transduction has advanced rapidly over the past
decade, evolving from low-efficiency proof-of-principle demonstrations to
devices that approach quantum-limited operation. This review has compared
optomechanical, electro-optic, and magneto-optic platforms encompassing conversion efficiency, added noise, cooperativity,
transduction bandwidth, and operating temperature. Two additional parameters
were introduced: the internal efficiency $\eta_{\mathrm{in}}$, which isolates
intrinsic conversion performance from port-coupling losses, and the magnon decay
rate $\kappa_m/2\pi$, which governs efficiency, bandwidth, cooperativity, and
thermal noise simultaneously in magneto-optic systems and has not been
consistently reported in prior comparative studies.

Optomechanical systems have demonstrated near-unity internal efficiency and
sub-quantum added noise, establishing them as the current benchmark for
high-fidelity quantum state transfer. Electro-optic systems offer the widest
bandwidths and the strongest integration prospects with superconducting
processors, with internal efficiencies approaching 100\% and recent experiments
demonstrating direct optical control of superconducting qubits. Magneto-optic
systems remain efficiency-limited by weak optomagnonic cooperativity, but their
inherent non-reciprocity and magnetic tunability give them a unique functional
niche that the other platforms cannot replicate.

A core finding of this review is that the three platforms are not competing
alternatives but complementary technologies, each best suited to a distinct
function within a heterogeneous quantum network. The fundamental tension among
efficiency, noise, bandwidth, and operating temperature cannot be resolved by
any single platform in its current form. Progress will require continued
advances in materials, nanofabrication, thermal engineering, and system-level
integration. Collectively, the field is converging toward the performance
levels needed for practical distributed quantum computing and large-scale
quantum communication networks.

\section*{Acknowledgements}
The authors thank Dr.\ Sajid Muhaimin Choudhury for guidance on the topic and
for encouraging research in quantum information science.
\section*{Author Contributions}

T.A.A.\ conceptualized the review structure and contributed the introduction,
added-noise formalism, optomechanical platform analysis, and performance metric
framework. J.S.I.\ contributed the quantum capacity and continuous-time quantum
capacity framework, transduction bandwidth and operating temperature metrics,
electro-optic platform analysis, and the cross-platform comparison.
K.Z.\ contributed the general transduction input-output formalism, transduction
efficiency and internal efficiency metrics, magneto-optic platform analysis,
and the conclusion. All three authors contributed to writing, revision, and
the final outlook and future direction section.
\bibliographystyle{unsrtnat}
\bibliography{references}

\end{document}